\documentclass[journal]{IEEEtran}
\usepackage[acronym]{glossaries}
\usepackage{rotating}
\usepackage{bbm}
\usepackage{float}
\usepackage{physics}
\usepackage{siunitx}
\usepackage[justification=centering]{caption}
\usepackage[table,xcdraw]{xcolor}
\usepackage[utf8]{inputenc}
\usepackage{listings}
\usepackage{subcaption}
\usepackage{tocloft}
\usepackage{indentfirst}
\usepackage{enumitem}
\usepackage{amsmath,amssymb,amsfonts,mathtools, nccmath}
\usepackage{filecontents,lipsum}
\usepackage{calrsfs}
\usepackage{verbatim}
\usepackage{amsthm,epsfig,epstopdf,titling,url}
\usepackage{graphics}
\usepackage{tabularray}

\usepackage{multicol}
\usepackage{multirow}
\usepackage{array}
    \newcolumntype{P}[1]{>{\centering\arraybackslash}p{#1}}
    \newcolumntype{M}[1]{>{\centering\arraybackslash}m{#1}}
\usepackage{cite}
\usepackage[font=scriptsize]{caption}

\DeclareMathAlphabet{\pazocal}{OMS}{zplm}{m}{n}

\DeclareMathOperator*{\argmax}{argmax}
\DeclareMathOperator*{\argmin}{argmin}
\def\acs@author@fnsymbol#1{}
\usepackage{cuted}
\newtheorem{theorem}{Theorem}

\newtheorem{lemma}{Lemma}
\newtheorem{remark}{Remark}
\theoremstyle{plain}

\ifCLASSINFOpdf
\else
\fi
\makeatletter
\def\thanks#1{\protected@xdef\@thanks{\@thanks
        \protect\footnotetext{#1}}}
\makeatother

\begin{document}
%
\title{Whittle Index Based User Association in Dense Millimeter Wave Networks}
%
%
%

\author{Mandar R. Nalavade,~
        Gaurav S. Kasbekar,~\IEEEmembership{Member,~IEEE,} 
        Vivek S. Borkar,~\IEEEmembership{Fellow,~IEEE}
\thanks{The authors are with the Department
of Electrical Engineering, Indian Institute of Technology (IIT) Bombay,
Mumbai, 400076, India. Their email addresses are 22d0531@iitb.ac.in, gskasbekar@ee.iitb.ac.in, and borkar@iitb.ac.in. The work of M.R. Nalavade and G.S. Kasbekar is supported in part by the project with code RD/0121-MEITY01-001.}
}
\date{}

\maketitle

\begin{abstract}
We address the problem of user association in a dense millimeter wave (mmWave) network, in which each arriving user brings a file containing a random number of packets and each time slot is divided into multiple mini-slots. This problem is an instance of the restless multi-armed bandit problem, and is provably hard to solve. Using a technique introduced by Whittle, we relax the hard per-stage constraint that each arriving user must be associated with exactly one mmWave base station (mBS) to a long-term constraint and then use the Lagrangian multiplier technique to convert the problem into an unconstrained problem. This decouples the process governing the system into separate Markov Decision Processes at different mBSs. We prove that the problem is Whittle indexable, present a scheme for computing the Whittle indices of different mBSs, and propose an association scheme under which, each arriving user is associated with the mBS with the smallest value of the Whittle index. Using extensive simulations, we show that the proposed Whittle index based scheme outperforms several user association schemes proposed in prior work in terms of various performance metrics such as average cost, delay, throughput, and Jain's fairness index. 
\end{abstract}


%
\IEEEpeerreviewmaketitle

\section{Introduction}
%
%
%
%
\IEEEPARstart{T}{he} demand for wireless data traffic is rapidly increasing, driven by factors such as the Internet of Things (IoT), streaming services, smart grids, vehicular networks, and immersive technologies such as AR/ VR (Augmented Reality/ Virtual Reality) \cite{jgandrews}, \cite{magiwal}, and it is pushing the boundaries of existing technologies. To meet the increasing demand for data traffic sent over the wireless medium, several new technologies are being used, such as Wi-Fi 6, 6E, and 7, massive Multiple Input Multiple Output (MIMO), millimeter wave (mmWave) communication, etc. In particular, for the fulfillment of the exponentially increasing data demand, the large amount of untapped spectrum available in the mmWave bands is being utilized. Although mmWave communication can provide extremely high data rates \cite{mcudak}, it poses several challenges including high propagation loss, sensitivity to blockage, need for directional communication, etc. \cite{mcudak}, \cite{niuy}. These challenges lead to several differences in the optimal deployment of base stations (BSs), user association policies, handover schemes, sizes and shapes of cells, etc. \cite{jgandrews}, \cite{magiwal} in mmWave communication relative to sub-6 GHz communication. Novel strategies need to be designed to overcome the above challenges in mmWave networks \cite{ksakag}. 

User association is the process that determines the BS which an arriving user joins. The process of user association is critical in mmWave networks, and the design of a user association policy involves various challenges such  as balancing the load across different BSs, achieving a high spectral efficiency and energy efficiency of the network, etc. \cite{dliu}. Numerous schemes for user association in sub-6 GHz as well as mmWave networks have been proposed in prior work \cite{ysun, rdong, akhalili, qye, kshen, hcui, gyehzhang, tzhou, ssobhigivi, yzhangksun, cmafliu, abmadam, mfeng, aalizadeh, gskpnugg, dbethanabh, yxusmao, buwiz, sligchu, dweifdu, cdandrea, aalizadehmvu, yzhangldai, hzhangshuang, sgoyal, bsoleimani, kkhawam, rliuqchen, cpanrliu, emmohamed, msanaade, hkhanaelgabli, sgaoztang, hbenye, rliumleegyu, rliugyu, zsigchu, sksinghvsb}, and are reviewed in Section \ref{section2_lit_survey}. An index based approach using an index other than the \emph{Whittle index} \cite{pwhittle} has been proposed to solve the user association problem in \cite{ysun}. However, to the best of our knowledge, with the exception of our prior work \cite{sksinghvsb}, the user association problem has not been addressed using the theory of Whittle index in the existing research literature. Also, the results in \cite{sksinghvsb} are limited to a simplified model. In this paper, we study the user association problem in a dense mmWave network using the theory of Whittle index. The setting considered in this paper is significantly more general compared to that in \cite{sksinghvsb}, as explained in Section \ref{section2_lit_survey}. The concept of Whittle index was first introduced in \cite{pwhittle} and has been successfully used to solve a variety of problems: scheduling while minimizing functions of the age of information \cite{vtripathi, mchenkwu, bsombabu}, real time multicast scheduling for wireless broadcasts \cite{vraghunathan}, scheduling stochastic arrivals \cite{yphsu}, dynamic multi-channel allocation \cite{kliuqzhao, jwangxren}, processor sharing \cite{vsbspatta}, opportunistic scheduling \cite{vsbgskspatta}, scheduling web crawlers \cite{keavra, kavravsb}, user association \cite{sksinghvsb}, etc.

In this paper, we consider a dense mmWave network. Time is divided into slots of equal durations. A user arrives with a file containing a random number of packets at the end of each time slot. Every time slot is further divided into a fixed number of mini-slots. In each mini-slot, a BS serves a packet with a fixed probability (serving rate). The serving rates of different BSs may be different. A cost (holding cost) is incurred in every time slot in which a packet is present with a user associated with a BS. The objective is to associate each arriving user with exactly one BS so as to minimize the long-run expected average cost at all the BSs in the network. This is an instance of the \emph{restless multi-armed bandit problem} and is provably hard to solve \cite{chpapadi}. Using the technique introduced by Whittle \cite{pwhittle}, we relax the hard per-stage constraint that each user must be associated with exactly one BS, to a long-term constraint, and then use the Lagrangian multiplier technique to convert the problem into an unconstrained problem. This decouples the process governing the system into separate Markov Decision Processes (MDPs) at different BSs. We prove that the problem is Whittle indexable \cite{pwhittle} and present a scheme for computing the Whittle indices of different BSs. Also, we propose an association scheme under which, each arriving user is associated with the BS with the smallest value of the Whittle index. Using extensive simulations, we show that the proposed Whittle index based scheme outperforms several user association schemes proposed in prior work in terms of various performance metrics such as average cost, delay, throughput, and Jain's fairness index \cite{rjaindchiu}. 

Note that, in general, it is difficult to prove the Whittle indexability of a restless multi-armed bandit problem. Our main technical contribution is that we provide a rigorous proof of the fact that the user association problem in mmWave networks is Whittle indexable. 

The rest of the paper is organized as follows. Section \ref{section2_lit_survey} provides a review of related prior research literature. Section \ref{section3} describes the system model and problem formulation. In Section \ref{Section4_SA}, we analyze the stability of the Markov chain according to which 
the process at a BS evolves. Section \ref{section5_DPE} provides the dynamic programming equation for the problem. Various structural properties of the value function of the MDP at a BS are proved in Section \ref{Section6_prop}. The threshold behavior of the optimal policy is analyzed in Section \ref{Section7_thres}. In Section \ref{Section8_wi}, it is proved that the problem is Whittle indexable. A scheme for the computation of Whittle indices is proposed in Section \ref{Section9_comp}. 
Several user association policies proposed in prior work, with which we compare our proposed scheme via simulations, are described in Section \ref{Section10_other}. We provide simulation results in Section \ref{Section11_sim}  and conclusions and directions for future research in Section \ref{Section12_con}.

\section{Related Work} \label{section2_lit_survey}
Surveys on user association strategies in heterogeneous cellular networks (HetNets) and mmWave networks are provided in \cite{hramaz} and \cite{attiahml}, respectively.

First, we provide a review of prior work on user association in wireless networks other than mmWave networks. In \cite{rdong}, a user association scheme for multi-cell, multi-user, full-dimension MIMO networks is proposed,  which has  the goal of maximizing network capacity. This is achieved with the help of a three-step Gaussian belief propagation (GaBP)-based distributed solver with low computational complexity \cite{rdong}. In \cite{akhalili}, the user association problem in heterogeneous networks is formulated as a non-convex mixed integer non-linear programming (MINLP) problem and it is solved by the application of majorization-minimization (MM) theory tools, aiming to maximize the data rates of small cell users. In \cite{qye}, the objective of user association while balancing the load in HetNets  is achieved by addressing a network-wide utility maximization problem. The work in \cite{kshen} proposes a pricing-based user association scheme for downlink MIMO cellular networks. This scheme assigns BSs to users to maximize the utility based on a pricing strategy. A virtual pricing theory based user association bargaining mechanism for multiple tiers of BSs and terminal users is introduced in \cite{hcui}. Its goal is to facilitate load balancing among the cells in heterogeneous small cell networks. In \cite{gyehzhang}, an algorithm is proposed for maximizing the energy efficiency in a two-tier heterogeneous network with small cells by iteratively addressing joint user association and power allocation. Similarly, \cite{tzhou} addresses the energy efficient user association problem in cellular HetNets. User association schemes for load balancing in HetNets  and  Non-Orthogonal Multiple Access (NOMA) based cellular HetNets are proposed in \cite{ssobhigivi} and \cite{yzhangksun}, respectively. A user association scheme that prioritizes energy efficiency through robust optimization is presented in \cite{cmafliu}. In \cite{abmadam}, efficient user association and sub-channel assignment in multi-cell multi-carrier NOMA based networks with the goal of maximizing the energy efficiency are accomplished  using deep neural network (DNN) models. A distributed joint interference nulling and user association scheme is proposed in \cite{mfeng} to maximize the sum rate of all users while considering the limitations on the degrees of freedom of BSs in MIMO HetNets. In \cite{aalizadeh}, a scheme is proposed for user association with load balancing in multi-tier HetNets, which employs a reinforcement learning multi-armed bandit (MAB) technique and incorporates centralized and semi-distributed online algorithms. In \cite{gskpnugg}, a stochastic dynamic programming based numerical approach is introduced for determining the optimal client-AP association in a small Wi-Fi network. In \cite{dbethanabh}, a simple decentralized user-centric association scheme, which utilizes the concept of Nash equilibrium and aims to maximize the throughput, is proposed. A centralized user association algorithm to achieve proportional fairness in massive MIMO wireless networks is introduced in \cite{yxusmao}. A learning algorithm, which utilizes the MAB technique for the intelligent association of secondary users with the available secondary BSs in 5G HetNets, is proposed in \cite{buwiz}. Unsupervised neural network learning algorithms  are designed in \cite{sligchu} to address the optimization problem of user association and power allocation in visible light communication (VLC) and radio frequency (RF) HetNets. The Sum Rate Maximization Matching (SRMM) algorithm is introduced in \cite{dweifdu} to address the problem of jointly optimizing user association and resource allocation in multi-cell NOMA networks, with the aim of maximizing the overall sum rate. In \cite{cdandrea},  a user association scheme is designed to maximize the sum rate of uplink transmissions of users of a scalable cell-free massive MIMO system. However, unlike our work, none of the above papers \cite{rdong, akhalili, qye, kshen, hcui, gyehzhang, tzhou, yzhangksun, ssobhigivi, cmafliu, abmadam, mfeng, aalizadeh, gskpnugg, dbethanabh, yxusmao, buwiz, sligchu, dweifdu, cdandrea} addresses the problem of user association in a dense mmWave network.

We now provide a review of prior work on user association strategies in mmWave networks. In \cite{aalizadehmvu}, load balancing is achieved by addressing a network-wide utility maximization problem in mmWave MIMO cellular networks. Load balancing  along with the minimization of the outage probability in mmWave networks is addressed in \cite{yzhangldai}. The challenge of joint user association in ultra-dense mmWave networks has been addressed using a Lagrangian dual decomposition based optimization framework in \cite{hzhangshuang}. The objective is to maximize both energy and spectral efficiency while accounting for load balancing constraints, BSs' energy harvesting capabilities, and ensuring user Quality of Service (QoS). An optimal and fair cell selection policy is devised in \cite{sgoyal}, taking into account the reallocation cost associated with potential handovers and effectively addressing the erratic characteristics of the mmWave channel. A clustering based user association algorithm is proposed for dense mmWave  femto-networks in \cite{bsoleimani}. In \cite{kkhawam}, a coordinated framework is devised to obtain a centralized approach for user association in 5G HetNets containing both macro cells and small cells, and utilizing both mmWave and sub-6 GHz technology. In \cite{rliuqchen}, the design of an association scheme for multi-band access in mmWave HetNets involves the use of a Markov approximation framework. In \cite{cpanrliu}, the proposed solution for joint user association and resource allocation in mmWave communication systems utilizes a neural network based algorithm. An online learning methodology, which utilizes a centralized multi-player multi-armed bandit (MP-MAB) formulation with load balancing across arms, and three distinct centralized MP-MAB algorithms are proposed in \cite{emmohamed}. A scalable and flexible  algorithm, which has low complexity, based on deep reinforcement learning is designed in \cite{msanaade} to optimize the network sum rate in dense mmWave networks. In \cite{hkhanaelgabli}, machine learning based algorithms are proposed to solve the problem of vehicle to cell association in mmWave networks with the aim of maximizing the time averaged rate per vehicular user. Another machine learning based algorithm with the goal of maximizing the overall downlink rate, while fulfilling certain QoS requirements for each user, is proposed for the context of ultra-dense mmWave communication networks in \cite{sgaoztang}. A user association problem with the objective of maximizing the number of served users in a dense mmWave network, with the fulfillment of certain QoS criteria that depend on the received signal qualities, is studied in \cite{hbenye}. A novel machine learning-driven user association strategy designed to facilitate multi-connectivity in mmWave networks is studied in \cite{rliumleegyu}. An efficient algorithm, which aims to maximize the throughput of enhanced mobile broadband (eMBB) users while satisfying the reliability constraints of ultra-reliable low-latency communications (URLLC) users, is introduced in \cite{rliugyu}. In \cite{zsigchu}, an innovative user association scheme is introduced for ultra-dense mmWave networks with the objective of maximizing the system throughput while ensuring the fulfillment of the QoS requirements of the users. However, unlike our paper, none of the above papers \cite{aalizadehmvu, yzhangldai, hzhangshuang, sgoyal, bsoleimani, kkhawam, rliuqchen, cpanrliu, emmohamed, msanaade, hkhanaelgabli, sgaoztang, hbenye, rliumleegyu, rliugyu, zsigchu} proposes an index based user association policy for mmWave networks. 

In \cite{ysun}, the challenge of user equipment to cell association in HetNets is addressed through the utilization of the restless multi-armed bandits model. An index based association policy is proposed, which aims to maximize the long-term system throughput. However, the index considered in \cite{ysun} is different from the Whittle index, which is the focus of this paper. Also, the work in \cite{ysun} is in the context of HetNets, whereas that in this paper is in the context of a dense mmWave network. 

The closest to this paper are our prior work \cite{vsbspatta} and \cite{sksinghvsb}, which propose Whittle index based schemes for the processor sharing problem and the problem of user association in a dense mmWave network, respectively. Although some of the results and proofs in this paper are similar to those in \cite{vsbspatta} and \cite{sksinghvsb}, the model in this paper significantly  differs from and generalizes those in \cite{vsbspatta} and \cite{sksinghvsb}. In particular, in the model in this paper, each time slot is divided into multiple mini-slots and there is a potential departure of a packet from a BS in each mini-slot; in contrast, in the models in \cite{vsbspatta} and \cite{sksinghvsb}, a time slot is not divided into mini-slots. Hence, the departure process in the model in this paper differs from and is more general than those in the models in \cite{vsbspatta} and \cite{sksinghvsb}. Also, in the model in this paper, each arriving user brings a file containing a random number of packets, and the maximum number of packets in the file is an arbitrarily large positive integer; in contrast, in the models in \cite{vsbspatta} and \cite{sksinghvsb}, each arriving user brings only one packet. Hence, the arrival process in the model in this paper is a significant generalization of those in \cite{vsbspatta} and \cite{sksinghvsb}. Finally, the work in  \cite{vsbspatta} is in the context of processor sharing, whereas that in this paper is in the context of user association in a dense mmWave network. Due to the above differences, the analysis in this paper differs significantly from, and generalizes, those in \cite{vsbspatta} and \cite{sksinghvsb}. 

\section{System Model And Problem Formulation} \label{section3}
\begin{figure} 
    \centering
    \includegraphics[width=\linewidth]{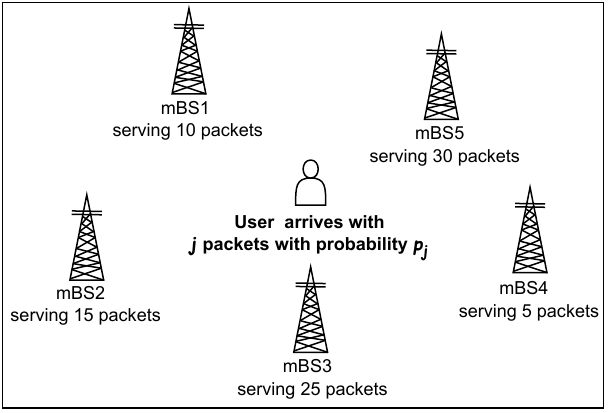}
    \captionsetup{justification   = justified, font=scriptsize}
    \caption{The figure shows an example of the system model with $K =  5$ mBSs, serving different numbers of packets. A user arrives with a file size of $j$ packets w.p. $p_j$.}
    \label{fig:sys_model}
\end{figure}

 Consider a dense mmWave wireless network comprising $K$ mmWave Base Stations (mBSs) catering to a small geographical area such as a bus stop, conference room, etc. Time is divided into slots of equal durations, and the slots are represented as $n \in \{0,1,2 \ldots \}$. At the end of every time slot $n$, a user arrives with a file of $j$ packets with probability (w.p.) $p_j$, where $j \in \{1, \ldots, M \}$ and no user arrives w.p. $p_0$, where $\sum_{j=0}^{M} p_j = 1$. Here, $M$ represents the maximum number of packets that a user can arrive with. When one or more users are already associated with a mBS, the introduction of a new user to that mBS implies that its $j$ packets must wait in a queue for service. Figure \ref{fig:sys_model} illustrates the system model for the case where there are $K = 5$ mBSs.

As the mBSs serve a small area, we assume that the serving rate $R_i$ of each packet with mBS $i \in \{1, 2, \ldots, K \}$ is the same for all the packets served by mBS $i$. The term $R_i$ is expressed in units of rate, i.e., bits per second. Consider a normalized serving rate of each packet by a mBS $i$, which is given by $r_i = \frac{R_i}{(\max_{\: l \in \left \{ 1,2, ..., K \right \}}R_l) + \zeta}$, where $\zeta$ is a small positive constant added to make $r_i \in (0,1)$. A time slot is further sub-divided into $L$ number of mini-slots, and we assume that whenever at least one packet is present with a user associated with mBS $i$, a single packet is served in a mini-slot with serving rate $r_i$. That is, a packet departs in a mini-slot w.p. $r_i$. Let $X_{n}^{i}$ be the total number of packets in the queues of all the users that are currently associated with mBS $i$, at the start of time slot $n$. Then the number of packet departures from the users associated with mBS $i$ in slot $n$ is $\min(X_n^i, \tilde{D}_n^i$), where $\tilde{D}_n^i$ is a binomial random variable with parameters $L$ and $r_i$. 
If $X_n^i$ is zero, then no departure will occur. If $X_n^i \geq 1$, then $d$ packets depart from mBS $i$ in time slot $n$ w.p. $P_{d}$, which is given by:
\begin{equation} \label{departure_prob}
P_{d}=
    \begin{cases}
        \binom{L}{d}  \left ( {r_i} \right )^d \left (  1-{r_i}\right )^{L-d}, & \text{if }d < \min(X_n^i,L),\\
        \sum_{l=d}^{L} \binom{L}{l}  \left ( {r_i} \right )^l \left (  1-{r_i}\right )^{L-l}, & \text{if } d=\min(X_n^i,L).
    \end{cases}
\end{equation}
The total number of packets at the users of mBS $i$ at the beginning of slot $n+1$ is given by:
\begin{equation} \label{update_state}
    X_{n+1}^{i} = (X_{n}^{i} - D_{n}^{i})^+ + \mu_{n}^{i} A_{n+1}, 
\end{equation}
where $x^+ := \max (0,x)$, $D_{n}^{i}$ is the number of departures of packets in time slot $n$, $A_{n+1}$ denotes the number of packets that arrive at the end of slot $n$, $\mu_{n}^{i}$ is the admission control variable, i.e., it indicates whether the arriving packets are admitted by mBS $i$ or not: 
\begin{equation*}
\mu_{n}^{i}=
    \begin{cases}
        1, & \text{if mBS } i \text{ admits the arriving user,}\\
        0, & \text{otherwise.}
    \end{cases}
\end{equation*}
Each arriving user needs to be associated with exactly  one mBS, which results in the constraint $\sum_{i=1}^{K} \mu_{n}^{i} = 1$. 

Suppose a holding cost of $C_i >0$ per packet is incurred in each time slot; then the total holding cost experienced at all the mBSs in the network in time slot $n$ is $\sum_{i=1}^{K} C_i X_{n}^{i}$. 

Our objective is to design a non-anticipating \cite{hordijk1983average} admission policy, which selects the variables $\mu_{n}^{i}$ so as to minimize the long-run expected average cost experienced at the mBSs in the network. The objective is as follows:
\begin{flalign} \label{primary_objective}
    \min \;\;\;\; &\limsup_{N \uparrow \infty} E \left [ \frac{1}{N} \sum_{n=0}^{N-1}\sum_{i=1}^{K}C_i\;X_{n}^{i}\right ], \notag\\
    &\text{s.t.} \;\; \sum_{i=1}^{K} \mu_{n}^{i} = 1, \;\;\forall n.
\end{flalign}
Minimization of the average cost in (\ref{primary_objective}) results in a low average delay experienced by packets in the network. 

The above constrained optimization problem is difficult to solve due to the exact per-stage constraint $\sum_{i=1}^{K} \mu_{n}^{i} = 1$ \cite{ysun}, \cite{chpapadi}. As in the scheme proposed by Whittle \cite{pwhittle}, we relax this constraint  to a time-averaged constraint,  which is given by:
\begin{equation} \label{relaxed_constraint}
\limsup_{N \uparrow \infty}  \frac{1}{N} \sum_{n=0}^{N-1}\sum_{i=1}^{K}E \left [\mu_{n}^{i}\right ] = 1.
\end{equation}
Using the Lagrangian multiplier approach \cite{borkarvivek}, we get an unconstrained problem with objective:
\begin{flalign} \label{unconstrained_problem}
    \min \;\;\;\; &\limsup_{N \uparrow \infty}  \frac{1}{N} \sum_{n=0}^{N-1}\sum_{i=1}^{K}E \left [\mathcal{F}_i(X_{n}^{i}, \mu_{n}^{i})\right ], \notag\\
    \text{where} \;\; &\mathcal{F}_i (x,u) = C_ix + \lambda \left ( 1 - \mu \right ),
\end{flalign}
and $\lambda$ is the Lagrangian multiplier. As we are considering a cost minimization problem, for the aforementioned form of the cost function, following the scheme proposed by Whittle \cite{pwhittle}, $\lambda$ is interpreted as a negative subsidy or a tax, i.e., if a mBS does not accept the arriving packets, then a tax is added to the cost incurred at that mBS.

When $\lambda$ is given, the above minimization problem (\ref{unconstrained_problem}) separates out into individual control problems at different mBSs. The control problem corresponding to each mBS takes on the form of a Markov Decision Process (MDP), in which the state is the total number of packets with the users associated with the mBS and the possible actions in a state are to accept or not to accept the arriving packets at the mBS.  This MDP is said to be \emph{Whittle indexable} if for all possible values of parameters $(C_i, r_i, p_j)$ 
and for every mBS, as $\lambda$ increases from $-\infty$ to $\infty$, the set of \emph{passive} states, i.e., those states in which the mBS does not accept the arriving packets, gradually diminishes in a monotonic fashion from encompassing the entire state space to eventually becoming an empty set. For every mBS, the \emph{Whittle index} for a state $x$ is the value of $\lambda$ for which the optimal controller of the MDP is indifferent between acceptance and non-acceptance of the arriving packets. Under the Whittle index based policy, in every time slot in which packets arrive, the mBS with the lowest  Whittle index  admits the incoming packets.

Without loss of generality, let us assume that $1 > r_1 \geq r_2 \geq ... \geq r_K > 0$; also, for the system at each mBS to be stable, we assume that $Lr_K > \sum_{j=0}^M j p_j$. In the next section, we provide a stability analysis of the system at each mBS. 

\begin{remark}
Note that the above system model is particularly well-suited for mmWave networks, in which dense deployments of mBSs are typically used. So $K$ is large and the above problem of user association is non-trivial since there are many possible mBSs with which an arriving user can possibly associate. However, all our results are also applicable to other wireless networks, e.g., sub-6 GHz wireless networks, in which BSs are densely deployed.      
\end{remark}

\section{Stability Analysis} \label{Section4_SA}
Under a particular control policy, the individual controlled MDP of each mBS gets converted into a Discrete Time Markov Chain (DTMC). For the induced DTMC, as the state is the number of packets, the state space $S$ is the set of all the non-negative integers, i.e., $S = \mathbb{W}$.
\begin{theorem}\label{theorem:therm1}
    If $Lr_K > \sum_{j=0}^M j p_j$, then the induced DTMC of each mBS $i$ is positive recurrent.
\end{theorem} 
Intuitively, the term $L r_K$ denotes the minimum average number of departures in a time slot from any mBS, whereas $\sum_{j=0}^{M} j  p_j$ denotes the average number of arrivals. So $L r_K > \sum_{j=0}^M j p_j$ indicates that the minimum average number of departures is strictly greater than the average number of arrivals. This condition ensures that, irrespective of the initial state, the controlled chain will inevitably return to state zero with a probability of $1$ in the future. This guarantees the stability (positive recurrence) of the controlled queue.
\begin{IEEEproof}
Consider a \emph{Lyapunov function} $\Psi(x) = x$, where $x$ denotes the total number of packets at the users of mBS $i$. The positive recurrence of the individual DTMC of mBS $i$ can be analyzed using Proposition 5.3 on p. 21 of \cite{sasmussen}. To prove the positive recurrence of the DTMC, we need to prove the following conditions:
\begin{flalign}
    \inf_{x \in S} \Psi(x) &> -\infty, \label{sta1}\\
    \sum_{y \in S} p_{xy} \Psi(y) &< \infty, \;\; \forall x \in S_0, \label{sta2} \\
    \Delta \Psi(x) =\sum_{y \in S} p_{xy} (\Psi(y)-\Psi(x)) &\leq -\epsilon, \; \forall x \notin S_0, \label{sta3}
\end{flalign}
where $S$ is the state space of the DTMC, $S_0$ is a finite set such that $S_0 \subset S$, $p_{xy}$ is the transition probability from state $x$ to state $y$ and $\Psi(x)$ is a Lyapunov function such that $\Psi : S \rightarrow \mathbb{R} $ and $\epsilon > 0$.

As $x \geq 0$, we can say that $\min \Psi(x) = \min x =  0$. Thus, $\inf_{x \in S} \Psi(x) > -\infty$. Hence, condition (\ref{sta1}) is satisfied.

Now, let $S_0 = \{0,1,\ldots,L-1\}$, where $L$ is the number of mini-slots, and let state $x \in S_0$. When the current state is $x \in S_0$, the highest possible next state that can be reached is $L+M-1$, where $M$ is the maximum number of arrivals in  a time slot. Similarly, the lowest possible next state it can go to is $0$. Also, all the transition probabilities are finite. Hence, $\sum_{x \in S}p_{xy}\Psi(x)$ is always finite. This proves  condition (\ref{sta2}).

Let $x \notin S_0$. From the assumption that $L r_K > \sum_{j=0}^M j p_j$, it follows that there exists a positive constant $\delta$ such that the difference $L r_K - \sum_{j=0}^M j p_j \geq \delta$. Also, choose $\epsilon$ such that $0 < \epsilon \leq \frac{\delta}{2}$. The drift in (\ref{sta3}) can also be expressed as $\Delta \Psi (x) = \mathbb{E} [\Psi(X_{t+1}) - \Psi(X_t)|X_t]$, where $X_{t+1}$ and $X_t$ are the total numbers of packets at the users of mBS $i$ at times $t+1$ and $t$, respectively. To prove the condition (\ref{sta3}), it is enough to show that: 
\begin{equation*}
    \Delta \Psi (x) = \mathbb{E} [\Psi(X_{t+1}) - \Psi(X_t)|X_t] \leq -\epsilon.
\end{equation*}
If $D$ and $A$ are the departure and arrival random variables, respectively, which are independent, then using equation (\ref{update_state}) and from the fact that the arrivals may or may not be accepted, the drift can be upper bounded as:
\begin{flalign*}
    \Delta \Psi (x) & \leq \mathbb{E}[-D|x] + \mathbb{E}[A|x] \leq -L r_K + \sum_{j=0}^M j p_j.
\end{flalign*}
However, by our assumption, $L r_K - \sum_{j=0}^M j p_j \geq \delta$ and $\epsilon \leq \frac{\delta}{2}$. Combining these inequalities, we get the condition (\ref{sta3}). This proves that the DTMC is positive recurrent. 
\end{IEEEproof}

\section{Dynamic Programming Equation} \label{section5_DPE}
As stated earlier, for a given value of $\lambda$, we can decouple the problem (\ref{unconstrained_problem}) into separate MDPs at the individual mBSs. Next, we focus on the MDP of mBS $i$. The proof of Whittle indexability of the decoupled problem corresponding to each mBS $i$ is the same; hence, for simplicity of notation, we drop the index $i$. Note that a maximum of `$\min (L, X_n)$' packets  depart in  slot $n$. Later in this section, we will prove that the Dynamic Programming Equation (DPE) for the separate MDP of a mBS can be expressed as (current state is $X_n = x$ and $\rho$ is a constant whose value will be specified later):

\begin{small}
\begin{equation*}
\begin{aligned} 
    V&(x)\\
    =& Cx - \rho + \min \biggl( \sum_{d=0}^{\min(L-1,x-1)} \binom{L}{d} \left ( r_i \right )^{d}\left (1 - r_i \right )^{L-d}\\ 
    &\biggl [ p_0 V(x-d) + \cdots + p_M V(x+M-d)  \biggl] \\ 
    &+ \sum_{d=\min(L,x)}^{L} \binom{L}{d} \left ( r_i \right )^{d} \left (1 - r_i \right )^{L-d}\\ 
    &\biggl [ p_0V(\max(0,x-L) + \cdots + p_M V(\max(0,x-L)+M) \biggl ];\\
    &\lambda + \sum_{d=0}^{\min(L-1,x-1)} \binom{L}{d} \left ( r_i \right )^{d}\left ( 1- r_i \right )^{L-d} V(x-d)\\
    &+ \sum_{d=\min(L,x)}^{L} \binom{L}{d} \left ( r_i \right )^{d}\left ( 1- r_i \right )^{L-d} V(\max(0,x-L) \biggl).
\end{aligned}
\end{equation*}
\end{small}
This can be simplified as:
\begin{flalign} \label{val_fun}
   V&(x)\notag\\
   =& Cx - \rho + \min \biggl( \sum_{d=0}^{\min(L-1,x-1)} \binom{L}{d} \left ( r_i \right )^{d}\left (1 - r_i \right )^{L-d}\notag\\
   &\biggl [ \sum_{q=0}^{M}p_q \biggl \{ V(x+q-d) - V(\max(0,x-L)+q) \biggl \} \biggl ]\notag\\
   &+ \sum_{q=0}^{M} p_q V(\max(0,x-L)+q); \lambda + V(\max(0,x-L)\notag\\
   &+ \sum_{d=0}^{\min(L-1,x-1)} \binom{L}{d} \left ( r_i \right )^{d}\left (1 - r_i \right )^{L-d}\notag\\
   &\biggl \{ V(x-d) - V(\max(0,x-L) \biggl \} \biggl ).   
\end{flalign}
The above result is obtained from the following identity about the sum of binomial coefficients:
\begin{flalign} \label{pro_identity} 
    1 =& \sum_{d=0}^{L} \binom{L}{d} \left (r_i\right )^d \left( 1-r_i\right)^{L-d}\notag\\ 
    =& \sum_{d=0}^{\min(L-1,x-1)} \binom{L}{d} \left (r_i\right )^d \left( 1-r_i\right)^{L-d}\notag\\
    &+ \sum_{d=\min(L,x)}^{L} \binom{L}{d} \left (r_i\right )^d \left( 1-r_i\right)^{L-d}.
\end{flalign}
In the rest of this section, we prove (\ref{val_fun}). The infinite horizon $\beta$-discounted cost for the controlled MDP with initial state $x$ (for $0 < \beta < 1)$, under any stationary policy $\pi$ is:
\begin{equation*}
    I^{\beta}(x, \pi) = E \left [  \sum_{n=0}^{\infty} \beta^n((1-\mu_n)\lambda + CX_n)|X_0 = x \right ].
\end{equation*}
The minimum over all stationary control policies is simply the value function of the above discounted problem and it is given by:
\begin{equation*}
    V^{\beta}(x) = \min_{\pi} I^{\beta}(x,\pi).
\end{equation*}
If $p_{.|.}(\mu)$ represents the transition probability of the controlled chain, then the value function can be described by the following DPE: 
\begin{equation*}
    V^{\beta}(x) = \min_{\mu} \left [ Cx + (1-\mu)\lambda + \beta\sum_{y}^{}p_{y|x}(\mu)V^{\beta}(y) \right ].
\end{equation*}
Consider $\bar{V}^{\beta}(.)=V^{\beta}(.)-V^{\beta}(0)$; using the above equation, it can also be expressed as:
\begin{equation} \label{beta_val_fun_bar}
\begin{split}
    \bar{V}^{\beta}(x) = &\min_{\mu} \biggl [ Cx + (1-\mu)\lambda - (1-\beta)V^{\beta}(0)\\
    &+ \beta\sum_{y}^{}p_{y|x}(\mu)\bar{V}^{\beta}(y) \biggl ].
\end{split}
\end{equation}
The value function $V$ and constant $\rho$ in (\ref{val_fun}) can be calculated with the help of the value function of the discounted problem given in (\ref{beta_val_fun_bar}) using the following lemma. 
\begin{lemma}  \label{lem:lemma1}
$V$ and $\rho$ specified in (\ref{val_fun}) can be calculated as: $\lim_{\beta \uparrow 1} \bar{V}^{\beta} = V$ and $\lim_{\beta \uparrow 1} (1-\beta) V^{\beta}(0) = \rho$. Moreover, $\rho$ is unique and equals the optimal cost $\rho(\lambda)$ in the RHS of (\ref{beta_val_fun_bar}) and the uniqueness of $V$ holds in states, which are positive recurrent under an optimal policy, subject to the additional constraint $V(0) = 0$. For a given state $x$, the $\argmin$ of the RHS of (\ref{val_fun}) gives the optimal choice of $\mu$. 
\end{lemma}
\begin{IEEEproof}
    This result can be proved similar to the  proof of Lemma 4 of \cite{vsbspatta}, with the change that the Lyapunov function specified in Section \ref{Section4_SA} is used instead of that specified in \cite{vsbspatta}. We omit the details of the proof for brevity. 
\end{IEEEproof}

\section{Structural Properties of Value Function} \label{Section6_prop}
Some characteristics of the value function that are required to prove the threshold nature of the optimal policy and subsequently the Whittle indexability are proved in this section.
\begin{lemma} \label{lem:lemma2}
    The function $V$ in (\ref{val_fun}) is non-decreasing in its argument.
\end{lemma}
\begin{IEEEproof}
    We use induction to prove the lemma. For $0 < q \leq s $, the DPE for the $s$-step finite horizon $\beta$-discounted problem is given by:
\begin{flalign} \label{finite_horizon_b_disc}
    V_{q}^{\beta}(x) =&\min_{\mu} \biggl [ \beta \sum_{j=0}^{M}  \biggl \{ \sum_{d=0}^{\min(x,L)} V_{q-1}^{\beta}(x-d+\mu j) P_d(x) \biggl \} p_j \notag\\ 
    &+c(x,\mu)\biggl ], 
\end{flalign}
with $V_{0}^{\beta}(x)=Cx, x \geq 0$ and $c(x,\mu) = Cx + (1-\mu)\lambda$. Here, $P_d(x)$ denotes the probability that $d$ packets depart when the current state is $x$ and is given by (\ref{departure_prob}).  $p_j$ denotes the probability of arrival of $j$ packets, where $j \in \{0, 1, \ldots, M\}$. Clearly, $V_{0}^{\beta}(x_1) > V_{0}^{\beta}(x_2)$ for all $x_1 > x_2$ and $x_1, x_2 \geq 0$. Assume that:
\begin{equation} \label{assumption1}
    V_{s-1}^{\beta}(x_1) > V_{s-1}^{\beta}(x_2), \;\;\;\;  \forall x_1 > x_2 \text{ and } x_1, x_2 \geq 0.
\end{equation}
There will be two cases based on the values of $x_2$ and $L$. For each of these cases, we need to show that $V_{s}^{\beta}(x_1) > V_{s}^{\beta}(x_2)$ for all $x_1 > x_2$ and $x_1, x_2 \geq  0$.

\textbf{Case I:} $0 < x_2 < L$ and $x_1 > x_2$. 
\begin{small}
\begin{equation*} 
\begin{split}
    V_{s}^{\beta}(x_1) =&\min_{\mu} \biggl [ c(x_1,\mu)\\ 
    &+\beta \sum_{j=0}^{M} \biggl \{ \sum_{d=0}^{\min(x_1,L)}  V_{s-1}^{\beta}(x_1-d+\mu j) P_d(x_1) \biggl \} p_j \biggl ]. 
\end{split}
\end{equation*}
\end{small}

Suppose the minimized values of $V_{s}^{\beta}(.)$ at $x_1$ and $x_2$ are obtained at $\mu_1$ and $\mu_2$, respectively. Then we can write: 
\begin{subequations} \label{v12}
    \begin{align}
    V_{s}^{\beta}(x_1) =&\beta \sum_{j=0}^{M} \biggl \{ \sum_{d=0}^{\min(x_1,L)}  V_{s-1}^{\beta}(x_1-d+\mu_1 j) P_d(x_1) \biggl \}p_j\notag\\
    &+Cx_1 + (1-\mu_1)\lambda,\\
    V_{s}^{\beta}(x_2) =&\beta \sum_{j=0}^{M} \biggl \{ \sum_{d=0}^{x_2}  V_{s-1}^{\beta}(x_2-d+\mu_2 j) P_d(x_2)\biggl \} p_j\notag\\
     &+ C x_2 + (1-\mu_2)\lambda.
\end{align}
\end{subequations}
For any $x_1 > x_2$, the following term is simplified as:
\begin{equation} \label{pro_split}
    \sum_{d=0}^{\min(x_1,L)}(.)P_d(x_1) = \sum_{d=0}^{x_2-1}(.)P_d(x_1) + \sum_{d=x_2}^{\min(x_1,L)}(.)P_d(x_1).
\end{equation}
The admissible controls $\mu_1$ and $\mu_2$ can take only binary values. Now, if both controls are equal, i.e., $\mu_1 = \mu_2 = \mu$, ($\mu$ can be either 0 or 1), then using (\ref{v12}) and (\ref{pro_split}), we get:
\begin{flalign} \label{v_diff}
    V&_{s}^{\beta}(x_1) - V_{s}^{\beta}(x_2)\notag\\
    =& C(x_1 - x_2)+ \beta \sum_{j=0}^{M}  \biggl [ \sum_{d=0}^{x_2-1} \biggl \{ V_{s-1}^{\beta}(x_1 - d + \mu j)\notag\\
    &- V_{s-1}^{\beta}(x_2 - d + \mu j) \biggl \}\Bar{P} \biggl ] p_j\notag\\
    &+ \beta \sum_{j=0}^{M}  \biggl [ \sum_{d=x_2}^{\min(x_1,L)} V_{s-1}^{\beta}(x_1 - d + \mu j)P_d(x_1)\notag\\
    &- V_{s-1}^{\beta}(\mu j) P_{x_2}(x_2) \biggl ] p_j, 
\end{flalign}
where $\Bar{P} = P_d(x_1)=P_d(x_2) = \binom{L}{d}(r_i)^d (1-r_i)^{L-d}$, when the number of departures is strictly less than the lowest current state. The first and second terms of (\ref{v_diff}) are always positive, since $x_1 > x_2$ and by (\ref{assumption1}). If the current state is $x \leq L$, then $P_{x}(x) = \sum_{d=x}^{L} \Bar{P}$. If $x_1 > L$, then (\ref{v_diff}) becomes:
\begin{flalign*}
    V&_{s}^{\beta}(x_1) - V_{s}^{\beta}(x_2)\\
    =& C(x_1 - x_2) + \beta \sum_{j=0}^{M}  \biggl [ \sum_{d=0}^{x_2-1} \biggl \{ V_{s-1}^{\beta}(x_1 - d + \mu j)\\
    &- V_{s-1}^{\beta}(x_2 - d + \mu j) \biggl \}\Bar{P} \biggl ] p_j\\
    &+ \beta \sum_{j=0}^{M}  \biggl [ \sum_{d=x_2}^{L} \biggl \{ V_{s-1}^{\beta}(x_1 - d + \mu j)  - V_{s-1}^{\beta}(\mu j) \biggl \} \Bar{P} \biggl ] p_j. 
\end{flalign*}
This is true because for a given range of the number of departures, $P_d(x_1) = \Bar{P}$. Using (\ref{assumption1}), we can say that $V_{s}^{\beta}(x_1) > V_{s}^{\beta}(x_2)$ for all $x_1 > x_2$. Also, if $x_1 \leq L$, then we can write (\ref{v_diff}) as:
\begin{flalign*}
    V&_{s}^{\beta}(x_1) - V_{s}^{\beta}(x_2)\\
    =& C(x_1 - x_2) + \beta \sum_{j=0}^{M}  \biggl [ \sum_{d=0}^{x_2-1} \biggl \{ V_{s-1}^{\beta}(x_1 - d + \mu j)\\
    &- V_{s-1}^{\beta}(x_2 - d + \mu j) \biggl \}\Bar{P} \biggl ] p_j\\
    &+ \beta \sum_{j=0}^{M}  \biggl [ \sum_{d=x_2}^{x_1-1} \biggl \{ V_{s-1}^{\beta}(x_1 - d + \mu j)  - V_{s-1}^{\beta}(\mu j) \biggl \} \Bar{P} \biggl ] p_j.
\end{flalign*}
Thus, we can say that $V_{s}^{\beta}(x_1) > V_{s}^{\beta}(x_2)$ for all $x_1 > x_2$. Now, let $\mu_1 = 1$ and $\mu_2 = 0$. We have already proved that:
\begin{flalign*}
    V_{s}^{\beta}(x_1)\biggl |_{\mu=1} > V_{s}^{\beta}(x_2)\biggl|_{\mu=1}.
\end{flalign*}
But if the minimum of $V_{s}^{\beta}(x_2)$ is achieved at $\mu=0$, we can write:
\begin{flalign*}
    V_{s}^{\beta}(x_2)\biggl|_{\mu=1} > V_{s}^{\beta}(x_2)\biggl|_{\mu=0}.
\end{flalign*}
Therefore, $V_{s}^{\beta}(x_1) > V_{s}^{\beta}(x_2)$. Similarly, we can prove the result for $\mu_1 = 0$ and $\mu_2=1$.

\textbf{Case II:} $x_2 \geq L$ and $x_1 > x_2$.\\
In this case, (\ref{v12}) can be written as:
\begin{flalign*}
    V_{s}^{\beta}(x_1) =&\beta \sum_{j=0}^{M} \biggl \{ \sum_{d=0}^{L}  V_{s-1}^{\beta}(x_1-d+\mu_1 j) \Bar{P}  \biggl \}  p_j\\
    &+ Cx_1 + (1-\mu_1)\lambda ,\\
    V_{s}^{\beta}(x_2) =&\beta \sum_{j=0}^{M} \biggl \{ \sum_{d=0}^{L}  V_{s-1}^{\beta}(x_2-d+\mu_2 j) \Bar{P} \biggl \} p_j\\
    &+ Cx_2 + (1-\mu_2)\lambda, 
\end{flalign*}
where $\Bar{P} = P_d(x_1) = P_d(x_2)$, when the number of departures is strictly less than the minimum of $L$ and the lowest current state. $\Bar{P}$ is the same as described in Case I. Similar to the proof in Case I, if both controls are equal, i.e., $\mu_1 = \mu_2 = \mu$, then from (\ref{assumption1}), we can say that $V_s^{\beta}(x_1) > V_s^{\beta}(x_2)$. The result can be proved for different values of $\mu_1$ and $\mu_2$ as in Case I. So for $x_1 > x_2$ and $x_1, x_2 \geq 0$, $V_{s}^{\beta}(x_1) > V_{s}^{\beta}(x_2)$. Applying limit as $s \uparrow \infty$, this equation also holds for the infinite horizon, i.e., $V^{\beta}(x_1) \geq V^{\beta}(x_2)$ and further applying limit as $\beta \uparrow 1$, we can say:
\begin{equation*}
    V(x_1) \geq V(x_2) \;\;\;\; \forall x_1 > x_2 \;\; \text{and} \;\; x_1\;,x_2\; \geq  0.
\end{equation*}
\end{IEEEproof}

\begin{lemma} \label{lem:lemma3}
$V$ in (\ref{val_fun}) has non-decreasing differences, i.e.:
\begin{equation*}
    w > 0, a > b \Rightarrow V(a+w) - V(a) \geq V(b+w) - V(b).
\end{equation*}
\end{lemma}
\begin{IEEEproof}
    Let us prove the convexity of $V$ first, which will imply non-decreasing differences. Consider the DPE for the $q$-step finite horizon $\beta$ discounted problem, for $0 < s \leq q$:
 \begin{equation*}
 \begin{split}
     V_{s}^{\beta}(x) =&\min_{\mu} \biggl [ c(x,\mu)\\
     &+ \beta\sum_{d=0}^{\min(x,L)} \sum_{j=0}^M V_{s-1}^{\beta}(x-d+\mu j) P_d(x) p_j \biggl ], 
\end{split}
 \end{equation*}
where $c(x,\mu) = Cx + (1-\mu)\lambda$ with $V_{0}^{\beta}(x)=Cx, x \geq 0$, which is convex. Here $P_d(x)$ denotes the probability of $d$ departures given the current state $x$ and $p_j$ denotes the probability of arrival of $j$ packets. The above equation can also be written as:
\begin{flalign} \label{modval}
    V_{s}^{\beta}(x) =&\min_{\mu} \biggl [ c(x,\mu)\notag\\ &+\beta\sum_{d=0}^{L} \sum_{j=0}^M V_{s-1}^{\beta}(x-x \wedge d +\mu j) P_d p_j \biggl ], 
\end{flalign}
where $x\wedge d = \min (x,d)$ and $P_d = \binom{L}{d} (r_i)^d (1-r_i)^{L-d}$, which is independent of $x$. We now relax the state space from the set of non-negative integers to non-negative real values, i.e., instead of considering discrete values of $x$ (total number of packets with the users of the mBS in a slot), we permit $x$ to assume any real value in $[0, \infty)$. The control space is also relaxed from $\{0,1\}$ to $[0,1]$. The departures $d$ and arrivals $j$ take only non-negative integer values.

We first show that $V_s^{\beta}(x)$ is convex $\forall s$, using an induction method. $V_0^{\beta}(x)=Cx$, which is convex. Assume that $V_{s-1}^{\beta}(x)$ is convex. Consider two points $x_1$ and $x_2$ with $x_1 > x_2$. Suppose the minima of $V_s^{\beta}(.)$ with states $x_1$ and $x_2$ are obtained at $\mu_1$ and $\mu_2$, respectively. Thus:
\begin{subequations} \label{v12_convex}
    \begin{align} 
     V_{s}^{\beta}(x_1) =&\beta\sum_{d=0}^{L} \sum_{j=0}^M V_{s-1}^{\beta}(x_1-x_1 \wedge d +\mu_1 j) P_d p_j\notag\\
    &+c(x_1,\mu_1), \label{18a}\\
    V_{s}^{\beta}(x_2) =&\beta\sum_{d=0}^{L} \sum_{j=0}^M V_{s-1}^{\beta}(x_2-x_2 \wedge d +\mu_2 j) P_d p_j \notag\\
    &+c(x_2,\mu_2). \label{18b}
\end{align}
\end{subequations}
Adding (\ref{18a}) and (\ref{18b}), we get:
\begin{flalign*}
    V&_{s}^{\beta}(x_1) + V_{s}^{\beta}(x_2)\\
    =& c(x_1,\mu_1) + \beta\sum_{d=0}^{L} \sum_{j=0}^M \biggl [ V_{s-1}^{\beta}(x_1-x_1 \wedge d +\mu_1 j)\\
    &+ V_{s-1}^{\beta}(x_2-x_2 \wedge d +\mu_2 j) \biggl ] P_d p_j + c(x_2,\mu_2).
\end{flalign*}
$c(x,\mu)$ is linear and, by our assumption, $V_{s-1}^{\beta}(.)$ is convex, hence:

\begin{flalign} \label{changing_equation}
    V&_{s}^{\beta}(x_1) + V_{s}^{\beta}(x_2) \notag\\
    \geq& 2 \biggl [ c \biggl ( \frac{x_1+x_2}{2}, \frac{\mu_1+\mu_2}{2} \biggl ) +  \beta\sum_{d=0}^{L} \sum_{j=0}^M  V_{s-1}^{\beta} \biggl (\frac{x_1+x_2}{2} \notag \\
    &- \frac{x_1 \wedge d+x_2 \wedge d}{2} + \frac{\mu_1+\mu_2}{2} j\biggl ) P_d p_j \biggl ].
\end{flalign}
Let us prove the following inequality for any $x_1 > x_2$:
\begin{equation} \label{inequality}
    \frac{x_1 \wedge d+x_2 \wedge d}{2} \leq \biggl ( \frac{x_1+x_2}{2} \biggl)\wedge d.
\end{equation}
Consider different cases, depending on the values of $x_1$ and $x_2$. 

\textbf{Case 1:} $x_2 \geq d \implies x_1 > d$ and $\frac{x_1+x_2}{2} > d$.\\
Then the above inequality holds as the LHS of (\ref{inequality}) becomes $d$, which is equal to the RHS of (\ref{inequality}).

\textbf{Case 2:} $x_1 \leq d \implies x_2 < d$ and $\frac{x_1+x_2}{2} < d$.\\
Then the above inequality holds as the LHS of (\ref{inequality}) becomes $\frac{x_1+x_2}{2}$, which is equal to the RHS of (\ref{inequality}). 

\textbf{Case 3:} $x_2 < d$, $x_1 \geq d$ and $\frac{x_1+x_2}{2} \geq d$. 
\begin{flalign*}
    \text{LHS of (\ref{inequality})} &=  \frac{d+x_2}{2} \text{  and  } \text{RHS of (\ref{inequality})} = d 
\end{flalign*}
As $d > x_2$, it yields LHS of (\ref{inequality}) $<$ RHS of (\ref{inequality}).

\textbf{Case 4:} $x_2 < d$, $x_1 \geq d$ and $\frac{x_1+x_2}{2} < d$. 
\begin{flalign*}
    \text{LHS of (\ref{inequality})} &=  \frac{d+x_2}{2} \text{  and  } \text{RHS of (\ref{inequality})} = \frac{x_1+x_2}{2}  
\end{flalign*}
As $d \leq x_1$, it yields LHS of (\ref{inequality}) $\leq$ RHS of (\ref{inequality}). This proves inequality (\ref{inequality}). Using (\ref{inequality}) and Lemma \ref{lem:lemma2}, we can say:
\begin{equation*}
\begin{split}
    V&_{s-1}^{\beta} \biggl (\frac{x_1+x_2}{2} - \frac{x_1 \wedge d+x_2 \wedge d}{2} + \frac{\mu_1+\mu_2}{2} j\biggl )\\
    &\geq V_{s-1}^{\beta} \biggl (\frac{x_1+x_2}{2} - \biggl ( \frac{x_1+x_2}{2} \biggl)\wedge d + \frac{\mu_1+\mu_2}{2} j\biggl ).
\end{split}
\end{equation*}
Therefore, (\ref{changing_equation}) can be written as:
\begin{flalign*}
    V&_{s}^{\beta}(x_1) + V_{s}^{\beta}(x_2) \notag \\
    \geq& 2 \biggl [ c \biggl ( \frac{x_1+x_2}{2}, \frac{\mu_1+\mu_2}{2} \biggl ) +  \beta\sum_{d=0}^{L} \sum_{j=0}^M  V_{s-1}^{\beta} \biggl (\frac{x_1+x_2}{2}\\
    &- \biggl ( \frac{x_1+x_2}{2} \biggl)\wedge d + \frac{\mu_1+\mu_2}{2} j\biggl ) P_d p_j \biggl ]\\
    \geq& 2 V_{s}^{\beta} \biggl ( \frac{x_1+x_2}{2}  \biggl ).
\end{flalign*}
Thus, $V_s^{\beta}(x)$ is convex in $x$. Applying limit as $s \uparrow \infty$, this assertion extends to an infinite horizon $\beta$-discounted cost scenario and then by taking limit as $\beta \uparrow 1$, we can show the convexity of $V(x)$. Convexity implies non-decreasing differences. Hence, $V$ exhibits non-decreasing differences. 

The result derived above holds true not only for a continuous control and state space but is also valid for a discrete control and state space. The rest of this proof justifies this claim. Let us rewrite (\ref{val_fun}) in the form of minimization of  a convex function. It can be expressed as:
\begin{flalign*}
    V(x) =& Cx - \rho + \lambda \\
    &+ \min_{\mu \in [0,1]} \left [ \sum_{d=0}^{L} \sum_{j=0}^M V(x-x \wedge d +\mu j) P_d p_j -\mu \lambda \right ] \\
    =&Cx - \rho + \lambda + \sum_{d=0}^{L} V(x-x \wedge d) P_d p_0 \\ &+\min_{\mu \in [0,1]} \left [ \sum_{d=0}^{L} \sum_{j=1}^M V(x-x \wedge d +\mu j) P_d p_j -\mu \lambda \right ]. 
\end{flalign*}

In the objective function being minimized, $P_d$ and $p_j$ are positive constants in $(0,1)$. $V(x)$ is a convex increasing function. The preceding objective function being minimized can be reformulated as follows:
\begin{equation*}
    G(x,\mu) :=F(x,\mu) -\mu \lambda, 
\end{equation*}
where $F(x,\mu) = \sum_{d=0}^{L} \sum_{j=1}^M V(x-x \wedge d +\mu j) P_d p_j$ is a jointly convex increasing function in $x$ and $\mu$. Suppose $G$ has a unique minimizer $\mu^* \in [0,1]$. (The following steps can be adapted to deal with the cases in which the minimizer is not unique.) The convexity of $F$ guarantees the existence of its right derivative, $F_{+}^{'}$, and left derivative, $F_{-}^{'}$, at all but at most countably many points. Additionally, these derivatives are monotonically increasing with $F_{+}^{'}(x,\mu) \geq F_{-}^{'}(x,\mu), \forall x \in \mathbb{R}^+$ and a fixed $\mu$. Then we must have:
\begin{equation*}
    \lambda \in \partial F(x,\mu^*) = [F_{-}^{'}(x,\mu^*),F_{+}^{'}(x,\mu^*)].
\end{equation*}
\textbf{Claim:} Consider a state $y < x$. If the minimum of $G(y,\mu)$ is achieved at $\mu_y \in (0,1$), then $\mu_y \geq \mu_x$.
\begin{IEEEproof}
$F$ is a jointly convex increasing function in $x$ and $\mu$. For states $x, y$ with $y < x$, the difference $F(x,\mu) - F(y,\mu)$, and hence the difference $G(x,\mu)-G(y,\mu)$, increases as $\mu$ increases. Now, consider an interval $\textrm{U} := (\mu_x, 1]$. From the fact that $\mu_x$ is the minimizer, we get: $G(x,\mu_x) \leq G(x,\mu)|_{\mu \in \textrm{U} }$. Thus, for any interval in $\textrm{U}$, say $[\mu_1,\mu_2]$ with $\mu_1 < \mu_2$:
\begin{flalign*}
    \frac{\left[F(x,\mu_2) - \lambda \mu_2 \right] - \left[F(x,\mu_1) - \lambda \mu_1 \right] }{\mu_2 - \mu_1} &\geq 0.
\end{flalign*}
If $y < x$, then $G(y,\mu) < G(x, \mu) \; \forall \mu \in [0,1]$ and if the above equation does not hold, then it violates the non-decreasing differences property. Therefore, the minimizer of $G(y,\mu)$ cannot be less than $\mu_x$, i.e., $\mu_y \geq \mu_x$. Alternatively, we can prove, for states $z > x$, if the minimizer of $G(z,\mu)$ is $\mu_z$, then $\mu_z \leq \mu_x$.
\end{IEEEproof}

Suppose for a given $\lambda$ and state $x$, there exists a state $a <x$ such that $a \in \mathbb{W}$, for which the minimizer of $G(a,\mu)$, i.e., $\mu^* \in (0,1)$, and the minimizer at state $a-1$ (if the state exists), is unity, i.e., $\mu^* = 1$. Similarly, there exists a state $b>x$ such that $b \in \mathbb{W}$, for which the minimizer of $G(b,\mu)$, i.e., $\mu^* \in (0,1)$ and the minimizer at $b+1$ is zero, i.e., $\mu^* = 0$. Over the states $[a,b]$, we will get a fractional value of the control. For such states, if we restrict the control to $\{0,1\}$, there is at most one state in $[a,b]$ that will act as a threshold. The minimum would have been obtained at $\mu^*=0$ if $F(x,0) \leq F(x,1)-\lambda$ and at $\mu^*=1$ otherwise. This means that the states below the threshold will choose the optimal control as unity and the states above the threshold (including it) will choose the optimal control as zero due to the convex increasing nature of $F$. This still maintains the threshold property of the optimal policy, which is discussed in Section \ref{Section7_thres}. Thus, the dynamics remain in the original paradigm of the state and control space.
\end{IEEEproof}

\section{Threshold Behavior of Optimal Policy} \label{Section7_thres}
Under the optimal policy, there is a state which acts as a threshold for acceptance and non-acceptance of the arrival, for a given $\lambda$, i.e. states below or equal to the threshold accept the arrival and states after the threshold reject the arrival. Specifically, the threshold depends on the parameter $\lambda$ and for each $\lambda$, the optimal threshold policy for each decoupled process can be derived. The threshold nature is proved by the following lemma.
\begin{lemma} \label{lem:lemma4}
The optimal policy is a threshold policy.    
\end{lemma}
\begin{IEEEproof}
Let $f(x) = E[V((x-D)^+ +A)] - E[V((x-D)^+)]$, where $x$ is the total number of packets with the users of mBS $i$, the random variable $D$ is the number of departures, with $D \sim Bin(L, r_i)$, the random variable $A$ is the number of arrivals and $x^+$ is $\max(0,x)$. By definition of threshold policy, it follows that an optimal policy is a threshold policy iff $f(x+z) \geq f(x)$ for $z>0$. Therefore, it suffices to show that $f(x+1)-f(x) \geq 0$.

\begin{small}
\begin{flalign*}
    E&[V(x-D)^+ +A)] \\
    =&\sum_{d=0}^{\min(L-1,x-1)} \binom{L}{d}\left( r_i \right )^d \left ( 1- r_i \right )^{L-d} \left[  \sum_{q=0}^{M} p_q V(x+q-d) \right ]\\
    &+ \sum_{d=\min(L,x)}^{L} \binom{L}{d}\left ( r_i \right )^d \left ( 1- r_i \right )^{L-d} \times \\
    &\left[ \sum_{q=0}^{M} p_q V(\max(0,x-L)+q) \right]
\end{flalign*}
\end{small}

Using (\ref{pro_identity}), the above equation can be simplified to the following:

\begin{flalign*}
    E&[V(x-D)^+ +A]\\
    =&\sum_{q=0}^{M} p_q V(\max(0,x-L)+q)\\
    &+ \sum_{d=0}^{\min(L-1,x-1)} \binom{L}{d}\left ( r_i \right )^d \left ( 1- r_i \right )^{L-d} \times\\
    &\left[ \sum_{q=0}^{M} p_q \biggl \{ V(x+q-d) - V(\max(0,x-L)+q) \biggl \} \right].
\end{flalign*}
Similarly:

\begin{small}
\begin{flalign*}
    E&[V(x-D)^+]\\
    =&V(\max(0,x-L)+ \sum_{d=0}^{\min(L-1,x-1)} \binom{L}{d}\left ( r_i \right )^d \left ( 1- r_i \right )^{L-d} \times \\
    &\biggl [  V(x-d) - V(\max(0,x-L))  \biggl ].
\end{flalign*}
\end{small}

The function $f(x)$ can be expressed as:
\begin{small}
\begin{flalign*}
   f&(x)\\
   =&E[V(x-D)^+ +A] - E[V(x-D)^+]\\
   =&\sum_{d=0}^{\min(L-1,x-1)} \binom{L}{d}\left ( r_i \right )^d \left ( 1- r_i \right )^{L-d} \times \\
   &\biggl \{  \left [ \sum_{q=0}^{M} p_q \{ V(x+q-d) - V(\max(0,x-L)+q) \} \right]\\    
   &- \{ V(x-d) - V(\max(0,x-L)) \} \biggl \}\\
   &+ \sum_{q=0}^{M} p_q V(\max(0,x-L)+q) - V(\max(0,x-L))\\
   =&\sum_{d=0}^{\min(L-1,x-1)} \binom{L}{d}\left ( r_i \right )^d \left ( 1- r_i \right )^{L-d} \times \\
   &\biggl [ p_0 V(x-d) + \cdots + p_M V(x+M-d) - V(x-d)\\
   &- p_0 V(\max(0,x-L)) - \cdots - p_M V(\max(0,x-L)+M)\\
   &+ V(\max(0,x-L)) \biggl ] + \biggl [ p_0 V(\max(0,x-L)) + \cdots\\
   &+ p_M V(\max(0,x-L)+M) - V(\max(0,x-L)) \biggl ] \\
      =&\sum_{d=0}^{\min(L-1,x-1)} \binom{L}{d}\left ( r_i \right )^d \left ( 1- r_i \right )^{L-d} \times \notag\\
   &\biggl [ p_1 V(x+1-d) + \cdots + p_M V(x+M-d)\notag\\
   &- (1-p_0) V(x-d) - p_1 V(\max(0,x-L)+1) - \cdots\notag\\
   &- p_M V(\max(0,x-L)+M)\notag\\
   &+ (1-p_0) V(\max(0,x-L)) \biggl ] + \biggl[ p_1 V(\max(0,x-L)+1)\notag\\ 
   &+ \cdots + p_M V(\max(0,x-L)+M)\notag\\
   &- (1-p_0) V(\max(0,x-L)) \biggl]\notag \\
   =&\sum_{d=0}^{\min(L-1,x-1)} \binom{L}{d}\left ( r_i \right )^d \left ( 1- r_i \right )^{L-d} \times \notag\\
   &\biggl [ \sum_{q=1}^{M} p_q \biggl \{ \biggl ( V(x+q-d) - V(x-d) \biggl )\notag\\
   &- \biggl ( V(\max(0,x-L)+q) - V(\max(0,x-L)) \biggl ) \biggl \} \biggl ]\notag\\
   &+ \sum_{q=1}^{M} p_q \biggl ( V(\max(0,x-L)+q) - V(\max(0,x-L)) \biggl ).\notag
\end{flalign*}
\end{small}

Similarly:

\begin{small}
\begin{flalign*}
   f&(x+1) \\
   =&\sum_{d=0}^{\min(L-1,x)} \binom{L}{d}\left ( r_i \right )^d \left ( 1- r_i \right )^{L-d} \times \\
   &\biggl [ \sum_{q=1}^{M} p_q \biggl \{ \biggl ( V(x+q+1-d) - V(x+1-d) \biggl )\\
   &- \biggl ( V(\max(0,x-L+1)+q)- V(\max(0,x-L+1)) \biggl )\biggl \} \biggl ]
\end{flalign*}
\begin{flalign*}
   &+ \sum_{q=1}^{M} p_q \biggl ( V(\max(0,x-L+1)+q)- V(\max(0,x-L+1)) \biggl ).
\end{flalign*}
\end{small}

There are two cases: the maximum number of departures, $\min(L,x)$, can be $x$ or $L$. Let us prove this lemma for both the cases. 

In Case $I$, where $\min(L,x+1) = x+1$, i.e., $\min(L-1,x)=x$ or $\min(L,x) = x$, $f(x)$ and $f(x+1)$ are:

\begin{small}
\begin{subequations} 
    \begin{align}
     f(x)=&\sum_{d=0}^{x-1} \binom{L}{d}\left ( r_i \right )^{d} \left ( 1- r_i \right )^{L-d} \notag\\
     &\biggl [ \sum_{q=1}^{M} p_q \biggl \{ \biggl ( V(x+q-d) - V(x-d) \biggl ) \notag\\ 
     &- \biggl ( V(q) - V(0) \biggl ) \biggl \} \biggl ] + \sum_{q=1}^{M} p_q\biggl(V(q) - V(0)\biggl),\label{fx_eq}
    \end{align}
    \begin{align}
    f&(x+1)\notag\\
    =& \sum_{d=0}^{x} \binom{L}{d}\left ( r_i \right )^{d} \left ( 1- r_i \right )^{L-d} \times \notag\\
    &\biggl [ \sum_{q=1}^{M} p_q \biggl \{ \biggl ( V(x+q+1-d) - V(x+1-d) \biggl )\notag\\
    &- \biggl ( V(q) - V(0) \biggl ) \biggl \} \biggl ] + \sum_{q=1}^{M} p_q \biggl (V(q) - V(0) \biggl )\notag\\
    =&\sum_{d=0}^{x-1} \binom{L}{d+1}\left ( r_i \right )^{d+1} \left ( 1- r_i \right )^{L-d-1} \times \notag\\
    &\biggl [ \sum_{q=1}^{M} p_q \biggl \{ \biggl ( V(x+q-d) - V(x-d) \biggl )\notag\\
    &-\biggl ( V(q) - V(0) \biggl ) \biggl \} \biggl ]\notag\\
    &+ (1-r_i)^L  \biggl [ \sum_{q=1}^{M} p_q \biggl \{ \biggl ( V(x+q+1) - V(x+1) \biggl )\notag\\
    &-\biggl ( V(q) - V(0) \biggl ) \biggl \} \biggl ] + \sum_{q=1}^{M} p_q \biggl (V(q) - V(0) \biggl ). \label{fx1_eq}
    \end{align}
\end{subequations}
\end{small}

Taking the difference between (\ref{fx1_eq}) and (\ref{fx_eq}), we get:

\begin{small}
\begin{flalign*}
  f&(x+1) -f(x) \\
  =& \sum_{d=0}^{x-1} \binom{L}{d+1}\left ( r_i \right )^{d+1} \left ( 1- r_i \right )^{L-d-1} \times \\
  &\biggl [ \sum_{q=1}^{M} p_q \biggl \{ \biggl ( V(x+q-d) - V(x-d) \biggl ) \\
  &-\biggl ( V(q) - V(0) \biggl ) \biggl \} \biggl ] + (1-r_i)^L  \\
  &\biggl [ \sum_{q=1}^{M} p_q \biggl \{ \biggl ( V(x+q+1) - V(x+1) \biggl ) -\biggl ( V(q) - V(0) \biggl ) \biggl \} \biggl ]\\
  &- \sum_{d=0}^{x-1} \binom{L}{d}\left ( r_i \right )^{d} \left ( 1- r_i \right )^{L-d} \times \\
  &\biggl [ \sum_{q=1}^{M} p_q \biggl \{ \biggl ( V(x+q-d) - V(x-d) \biggl ) - \biggl ( V(q) - V(0) \biggl ) \biggl \} \biggl ] 
\end{flalign*}
\begin{flalign*}
  f&(x+1) -f(x) \\
  =& (1-r_i)^L  \biggl [ \sum_{q=1}^{M} p_q \biggl \{ \biggl ( V(x+q+1) - V(x+1) \biggl )\\
  &- \biggl ( V(q) - V(0) \biggl ) \biggl \} \biggl ] \\
  &+ \sum_{d=0}^{x-1} \binom{L}{d}\left ( r_i \right )^{d} \left ( 1- r_i \right )^{L-d} \left [ \frac{L-d}{d+1} \frac{r_i}{1-r_i} - 1 \right] \times \\
  &\biggl [ \sum_{q=1}^{M} p_q \biggl \{ \biggl ( V(x+q-d) - V(x-d) \biggl )- \biggl ( V(q) - V(0) \biggl ) \biggl \} \biggl ].
\end{flalign*}
\end{small}

Let $\mathcal{B}(x) = \sum_{q=1}^{M} p_q(V(x+q)-V(x))$. From Lemma \ref{lem:lemma3}, $\mathcal{B}$ is a non-decreasing function. Thus:

\begin{small}
\begin{flalign*}
   f&(x+1) -f(x) \\
   =&\sum_{d=0}^{x-1} \binom{L}{d}(r_i)^d (1-r_i)^{L-d}\left [ \frac{L-d}{d+1} \frac{r_i}{1-r_i} -1 \right ] \times \\
   &\biggl [ \mathcal{B}(x-d) - \mathcal{B}(0) \biggl ] + (1-r_i)^{L} \biggl [ \mathcal{B}(x+1) - \mathcal{B}(0) \biggl ]\\
   =&\binom{L}{x-1} r_i^{x-1} (1-r_i)^{L-x+1} \left [ \frac{L-x+1}{x} \frac{r_i}{1-r_i} -1 \right ] \times \\
   &\biggl [ \mathcal{B}(1) - \mathcal{B}(0) \biggl ] + (1-r_i)^{L} \biggl [ \mathcal{B}(x+1) - \mathcal{B}(0) \biggl ]\\
   &+ \sum_{d=0}^{x-2} \binom{L}{d}(r_i)^d (1-r_i)^{L-d}\left [ \frac{L-d}{d+1} \frac{r_i}{1-r_i} -1 \right ] \times \\
   &\biggl [ \mathcal{B}(x-d) - \mathcal{B}(0) \biggl ]\\
   =&\binom{L}{x-1} r_i^{x-1} (1-r_i)^{L-x+1} \left [ \frac{L-x+1}{x} \frac{r_i}{1-r_i} -1 \right ] \times \\
   &\biggl [ \mathcal{B}(1) - \mathcal{B}(0) \biggl ] + (1-r_i)^{L} \biggl [ \mathcal{B}(x+1) - \mathcal{B}(0) \biggl ]\\
   &+ \biggl \{ \sum_{d=0}^{x-2} \binom{L}{d}(r_i)^d (1-r_i)^{L-d}\left [ \frac{L-d}{d+1} \frac{r_i}{1-r_i} \right ]\\
   &- \sum_{d=0}^{x-2} \binom{L}{d}(r_i)^d (1-r_i)^{L-d} \biggl \} \biggl [ \mathcal{B}(x-d) - \mathcal{B}(0) \biggl ]\\
    =&\binom{L}{x-1} r_i^{x-1} (1-r_i)^{L-x+1} \left [ \frac{L-x+1}{x} \frac{r_i}{1-r_i} -1 \right ] \times \\
   &\biggl [ \mathcal{B}(1) - \mathcal{B}(0) \biggl ] + (1-r_i)^{L} \biggl [ \mathcal{B}(x+1) - \mathcal{B}(0) \biggl ]\\
   &+ \biggl \{ \sum_{d=0}^{x-2} \biggl ( \binom{L}{d+1}(r_i)^{d+1} (1-r_i)^{L-d-1}\\
   &- \binom{L}{d}(r_i)^d (1-r_i)^{L-d} \biggl) \biggl \} \biggl [ \mathcal{B}(x-d) - \mathcal{B}(0) \biggl ]\\
   =&\biggl \{\sum_{d=0}^{x-2} \binom{L}{d}(r_i)^d (1-r_i)^{L-d} \biggl [ \mathcal{B}(x-d) - \mathcal{B}(0) \biggl ]\\
   &-\binom{L}{x-1} r_i^{x-1} (1-r_i)^{L-x+1} \left [ 1- \frac{L-x+1}{x} \frac{r_i}{1-r_i} \right ] \times \\
   &\biggl [ \mathcal{B}(1) - \mathcal{B}(0) \biggl ]\biggl \} + \biggl \{ (1-r_i)^{L} \biggl [ \mathcal{B}(x+1) - \mathcal{B}(0) \biggl ] \\
   &+ \sum_{d=0}^{x-2}  \binom{L}{d+1}(r_i)^{d+1} (1-r_i)^{L-d-1}  \biggl [ \mathcal{B}(x-d) - \mathcal{B}(0) \biggl ] \biggl \}
\end{flalign*}
\end{small}
\begin{small}
\begin{flalign*}
   f&(x+1) -f(x) \\
      =&\sum_{d=1}^{x-1}  \binom{L}{d}(r_i)^{d} (1-r_i)^{L-d}  \biggl [ \mathcal{B}(x-d+1) - \mathcal{B}(0) \biggl ]\\
   &- \sum_{d=0}^{x-1} \binom{L}{d}(r_i)^d (1-r_i)^{L-d} \biggl [ \mathcal{B}(x-d) - \mathcal{B}(0) \biggl ]\\
   &+ \binom{L}{x-1} r_i^{x-1} (1-r_i)^{L-x+1} \frac{L-x+1}{x} \frac{r_i}{1-r_i} \times \\
   &\biggl [ \mathcal{B}(1) - \mathcal{B}(0) \biggl ] + (1-r_i)^{L} \biggl [ \mathcal{B}(x+1) - \mathcal{B}(0) \biggl ]\\
   =&\sum_{d=0}^{x-1}  \binom{L}{d}(r_i)^{d} (1-r_i)^{L-d} \biggl [ \mathcal{B}(x-d+1) - \mathcal{B}(x-d) \biggl ] \\ 
   &+ \frac{L-x+1}{x} \binom{L}{x-1} r_i^{x} (1-r_i)^{L-x} \biggl [ \mathcal{B}(1) - \mathcal{B}(0) \biggl ] \geq 0.
\end{flalign*}
\end{small}

Note that using Lemma \ref{lem:lemma3}, $\mathcal{B}(x-d+1) - \mathcal{B}(x-d)$ and $\mathcal{B}(1)-\mathcal{B}(0)$ are always non-negative for any value of $d$. Thus, the difference $f(x+1)-f(x)$ is always non-negative, which shows that the optimal policy is a threshold policy. 

In Case II, where $\min(L,x+1)=L$ and $\min(L,x)=L$, we get:
\begin{small}
\begin{subequations} \label{case2}
    \begin{align} 
     f(x) =&\sum_{d=0}^{L-1}  \binom{L}{d}(r_i)^{d} (1-r_i)^{L-d} \times \notag\\
     &\biggl [ \sum_{q=1}^{M} p_q \biggl \{ \biggl ( V(x+q-d) - V(x-d) \biggl )\notag\\
     &- \biggl ( V(x-L+q) - V(x-L) \biggl  )\biggl \} \biggl ]\notag\\
     &+ \sum_{q=1}^{M} p_q \biggl ( V(x-L+q) - V(x-L) \biggl ), \label{22a}\\
     f(x+1) =&\sum_{d=0}^{L-1} \binom{L}{d}(r_i)^{d} (1-r_i)^{L-d} \times \notag\\
     &\biggl [ \sum_{q=1}^{M} p_q \biggl \{ \biggl ( V(x+q+1-d) - V(x+1-d) \biggl )\notag\\
     &- \biggl ( V(x+1-L+q) - V(x+1-L) \biggl  )\biggl \} \biggl ]\notag\\
     &+ \sum_{q=1}^{M} p_q \biggl ( V(x+1-L+q) -V(x+1-L) \biggl ). \label{22b}
\end{align}
\end{subequations}
\end{small}

\noindent Taking the difference between (\ref{22b}) and (\ref{22a}), we get:
\begin{flalign*}
  f&(x+1)-f(x)\\
  =&\sum_{d=0}^{L-1} \binom{L}{d}(r_i)^{d} (1-r_i)^{L-d} \times \\
   &\biggl [ \sum_{q=1}^{M} p_q \biggl \{ \biggl ( V(x+q+1-d) - V(x+q-d) \biggl ) \\
   &- \biggl ( V(x+1-d) - V(x-d) \biggl  )\biggl \} \biggl ] \\
   &+ \sum_{q=1}^{M} p_q \biggl [ \biggl ( V(x+1-L+q)- V(x-L+q)\biggl )
\end{flalign*}
\begin{flalign*}
  &- \biggl( V(x+1-L) - V(x-L)\biggl) \biggl ] r_i^L.
\end{flalign*}
Using Lemma \ref{lem:lemma3}, $V(x+q+1-d) - V(x+q-d)$ is always greater than $V(x+1-d) - V(x-d)$ for every value of $q \geq 1$. Similarly, $V(x+1-L+q)- V(x-L+q)$ is always greater than $V(x+1-L) - V(x-L)$. Thus, the difference $f(x+1)-f(x)$ is always positive. Hence, the optimal policy is a threshold policy in both cases. 
\end{IEEEproof}

\begin{lemma} \label{lem:lemma5}
If the stationary distribution of the induced DTMC under the threshold policy with threshold $t$ is $v_t$, then $\sum_{q=0}^{t}v_t(q)$ is an increasing function of $t$.
\end{lemma}
\begin{IEEEproof}
The result can be proved using Lemma \ref{lem:lemma4} similar to the proof of Lemma 8 on p. 11 of \cite{vsbspatta}. We omit the details of the proof for brevity. 
\end{IEEEproof}

\section{Whittle Indexability} \label{Section8_wi}
Whittle indexability means that as $\lambda$ decreases from $\infty$ to $-\infty$, the set of passive states steadily expands from the empty set to the entire state space. To prove Whittle indexability, we need some more lemmas.
\begin{lemma} \label{lem:lemma6}
    If $w : \mathbb{R} \times \mathbb{N} \rightarrow \mathbb{R}$ is submodular, i.e., $\forall \lambda_2 < \lambda_1$ and $x_2 < x_1$:
    \begin{equation*}
        w(\lambda_1, x_2) + w(\lambda_2, x_1) \geq w(\lambda_1, x_1) + w(\lambda_2, x_2),
    \end{equation*}
    and $x(\lambda) \coloneqq \inf \{x^* : w(\lambda, x^*) \leq w(\lambda, x) \; \forall x\}$, then $x(\lambda)$ is a non-decreasing function of $\lambda$.
\end{lemma}
\begin{IEEEproof}
    The proof is provided in Section 10.2 of \cite{rksundaram}.
\end{IEEEproof}
\begin{lemma} \label{lem:lemma7}
    Under the tax $\lambda$ and the threshold policy $\mu$, with threshold $k$, if the stationary average cost is:
    \begin{equation*}
        w(\lambda, k) = C \sum_{j=0}^{\infty} j \mu_k (j) + \lambda \sum_{j=k+1}^{\infty} \mu_k(j),
    \end{equation*}
    then $w$ is submodular.
\end{lemma}
\begin{IEEEproof}
    To prove this, we need to show that $\forall \lambda_2 < \lambda_1$ and $k_2 < k_1$,
\begin{equation*}
    w(\lambda_1, k_2) + w(\lambda_2, k_1) \geq w(\lambda_1, k_1) + w(\lambda_2, k_2).
\end{equation*}
The above inequality reduces to:
\begin{equation*}
\begin{split}
    \lambda_1 &\sum_{j=k_2 +1}^{\infty} \mu_{k_2}(j) + \lambda_2 \sum_{j=k_1 +1}^{\infty} \mu_{k_1}(j) \\ &\geq \lambda_1 \sum_{j=k_1 +1}^{\infty} \mu_{k_1}(j) + \lambda_2 \sum_{j=k_2 +1}^{\infty} \mu_{k_2}(j).
\end{split}
\end{equation*}
As $\lambda_1 > \lambda_2$, the above inequality further reduces to:
\begin{flalign*}
    \sum_{j=k_2 +1}^{\infty} \mu_{k_2}(j) &\geq \sum_{j=k_1 +1}^{\infty} \mu_{k_1}(j),\\
    \sum_{j=0}^{\infty} \mu_{k_2}(j) - \sum_{j=0}^{k_2} \mu_{k_2}(j) &\geq \sum_{j=0}^{\infty} \mu_{k_1}(j) - \sum_{j=0}^{k_1} \mu_{k_1}(j),\\
    \sum_{j=0}^{k_1} \mu_{k_1}(j) &\geq \sum_{j=0}^{k_2} \mu_{k_2}(j).
\end{flalign*}
The last inequality follows from Lemma \ref{lem:lemma5}. Thus, $w$ is submodular.
\end{IEEEproof}

\begin{theorem}
    The problem is Whittle indexable.
\end{theorem}
\begin{IEEEproof}
    For any stationary policy under the unichain property \cite{sheldonross}, there exists a unique stationary distribution. Denoting the unique stationary distribution by $v$ and the set of states where the mBS does not accept the incoming packets by $\emph{Z}$, if an arrival occurs under any stationary policy $\pi$, for a specified $\lambda$, the problem's optimal average cost can be written as:
\begin{equation*}
    \rho(\lambda) = \inf_{\pi} \left \{ C\sum_{j}^{} jv(j) + \lambda\sum_{j \in \emph{Z}}^{}v(j) \right \} \coloneqq w(\lambda, k(\lambda)).
\end{equation*}
Using Lemma \ref{lem:lemma7}, we can say that $w$ is submodular; also, from Lemma \ref{lem:lemma6}, as $\lambda$ increases, the threshold $k(\lambda)$ also increases. Under the threshold stationary policy, the form of the set of passive states will be $[k(\lambda), \infty)$ and therefore, we can say that with an increase in $\lambda$ from $-\infty$ to $\infty$, the set of passive states steadily diminishes from encompassing the entire state space to ultimately becoming an empty set. This implies that the problem is Whittle indexable.
\end{IEEEproof}

\section{Whittle Index Computation}\label{Section9_comp}
A recursive approach is used to compute the Whittle index. For a given state $x$, the parameter $\lambda$ is updated as follows:
\begin{small}
\begin{flalign} \label{update_lambda}
    \lambda_{t+1} =& \lambda_t + \gamma \biggl ( \sum_{i}^{}p_a(i|x)V_{\lambda_t}(i) - \sum_{i}^{}p_b(i|x)V_{\lambda_t}(i) - \lambda_t \biggl ),\notag\\
    &t \geq 0,
\end{flalign}
\end{small}

\noindent where $\gamma$ is a small positive step size, $p_a(.|.)$ is the transition probability when the mBS admits an arrival in the current slot, $V_{\lambda_t}(.)$ is the value function of the DPE (\ref{val_fun}) at $\lambda_t$ and $p_b(.|.)$ is the transition probability when the mBS does not accept any arrival. Equation (\ref{update_lambda}) shows that the parameter $\lambda$ converges to the value at which the mBS is indifferent between acceptance and non-acceptance of the arrival, and it decreases the difference between the arguments in the minimization function given in (\ref{val_fun}). The equation for $V_{\lambda}$ defines a linear system, allowing for iterative solution using the current $\lambda$ value at each step. To obtain the solutions, solve the system of equations specified below for $V = V_{\lambda_t}$ and $\rho = \rho(\lambda_t)$ using $\lambda = \lambda_t$;
\begin{flalign*}
   V(y) &= Cy - \rho + \sum_{z}^{} p_a(z|y) V(z), \;\;\;\;\;\;\;\; y \leq x,\\
   V(y) &= Cy + \lambda - \rho + \sum_{z}^{} p_b(z|y) V(z), \;\; y > x,\\
   V(0) &= 0.
\end{flalign*}
The Whittle index for a fixed state $x$ is the value reached after convergence of the iteration (\ref{update_lambda}), which is executed for a large number of states to mitigate  the computational cost, and later, interpolation is employed to calculate the Whittle indices for the remaining states. During every time slot, the mBS that has the lowest Whittle index is selected to accept the arrivals.

\section{Other Existing Policies}\label{Section10_other}
In Section \ref{Section11_sim}, we compare the performance of the proposed Whittle index based association policy with those of  different association policies proposed in prior work, which are described in \cite{vkgupta}. We briefly describe these policies in this section.   
\subsection{Random Policy}
In a time slot, if a user arrives with some packets, then the user is assigned to one of the mBSs, which is selected uniformly at random. 
\subsection{Load Based Policy}
In a time slot, if a user with packets arrives, the mBS with the lowest total number of packets with associated users at the start of the time slot is chosen, and the incoming user is associated with that mBS. Ties are broken at random. That is, in time slot $n$, mBS $\argmin_{i \in \{1,2,\ldots,K\}} X_n^i$ accepts the arrival (if any) in the slot. 
\subsection{SNR Based Policy}
In a time slot, upon arrival of a user with packets, the mBS with the highest data rate is chosen, and the incoming user is associated with that  mBS. Ties are broken at random. That is, in time slot $n$, mBS $\argmax_{i \in \{1,2,\ldots,K\}} r_i$, accepts the arrival (if any) in the slot. 
\subsection{Throughput Based Policy}
In a time slot, when a user with packets arrives, the mBS with the highest throughput is chosen, and the incoming user is associated with that mBS. Ties are broken at random. That is, in time slot $n$, mBS $\argmax_{i \in \{1,2,\ldots,K\}} \frac{r_i}{X_n^i + 1}$, accepts the arrival (if any) in the slot. 
\subsection{Mixed Policy}
In a time slot, upon arrival of a user with packets, the mBS chosen for association of that user is the one with the maximum weighted sum of data rate and a positive scalar multiplied by the throughput upon association. Ties are broken at random. That is, in time slot $n$, mBS $\argmax_{i \in \{1,2,\ldots,K\}} \left(0.2*r_i + \frac{r_i}{X_n^i + 1} \right)$ accepts the arrival (if any) in the slot. The value $0.2$ is chosen since the performance of the policy has been empirically found to be good for that value in \cite{gskjkuri}.

\section{Simulations}\label{Section11_sim}
In this section, we compare the performance of the proposed Whittle index based user association policy with those of the policies described in Section \ref{Section10_other} via simulations performed using MATLAB. The performances of the various policies are compared in terms of the long-run average cost per slot,  average delay, average throughput, and Jain's Fairness Index (JFI) \cite{rjaindchiu}. The average delay is  the average number of mini-slots required for a user to depart from the system after arrival. Let $Q_i$ and $D_i$ be the number of packets that the $i^{th}$ user arrives with and the delay of the user, respectively. Then the throughput is calculated as $\frac{Q_i}{D_i}$. The average throughput is the average of the throughput of all the users. If $\top_i$ is the throughput of the $i^{th}$ user and the total number of users is $Y$, then the JFI is defined as: $\frac{\left ( \sum_{j=1}^{Y} \top_j \right )^2}{Y \sum_{j=1}^{Y}  \top_j^2}$. The JFI lies between $0$ and $1$ and increases with the degree of fairness of the distribution of the throughput of different users \cite{rjaindchiu}. 
\begin{figure} [h]
    \centering
    \begin{subfigure}[b]{0.24\textwidth}
        \includegraphics[width=\textwidth]{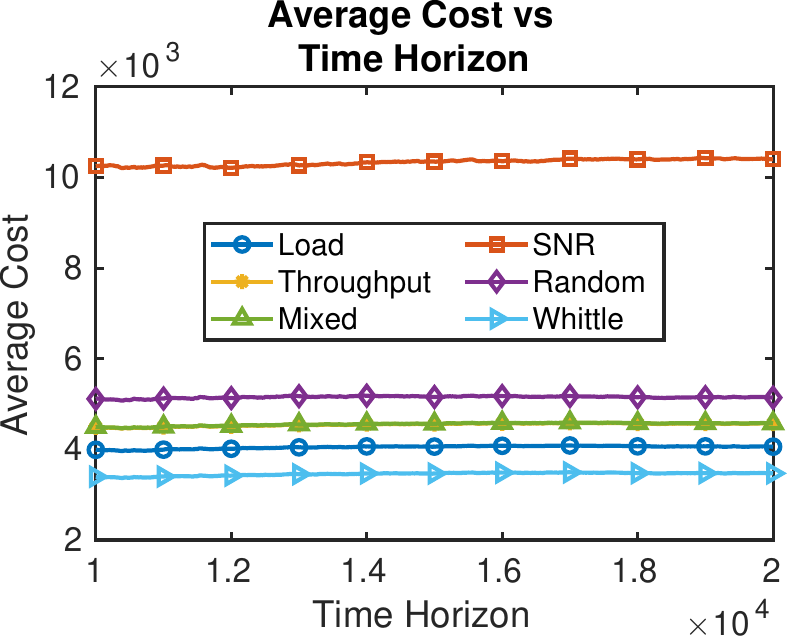}
        \caption{ }
        \label{fig:2a}
    \end{subfigure}
    \begin{subfigure}[b]{0.235\textwidth}
        \includegraphics[width=\textwidth]{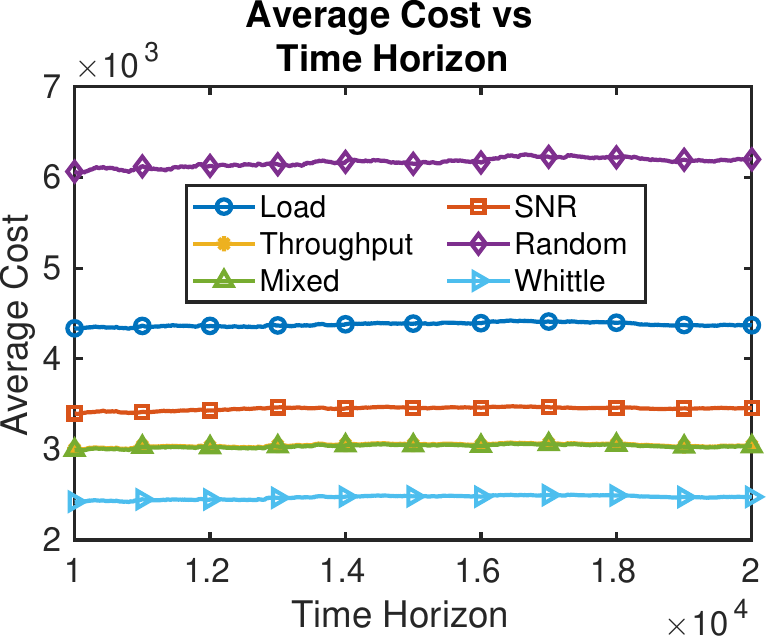}
        \caption{ }
        \label{fig:2b}
    \end{subfigure}
    \captionsetup{justification   = justified, font=scriptsize}
    \caption{The figures compare the average costs achieved under the six association policies. The following parameter values are used for both the plots: $K=5$, $M=100$, $L=20$, $r=[0.78,0.65,0.56,0.50,0.45]$, $p_0 = 0.6$, $p_j = 0.004,\; \forall j \in \{1,2,\ldots,M\}$. Also, for figure (a) (respectively, (b)), the cost vector $C=[95,75,58,40,32]$, (respectively, $C=[32,40,58,75,95]$) is used.}
    \label{fig:2}
\end{figure} 

\begin{figure} [h]
    \centering
    \begin{subfigure}[b]{0.24\textwidth}
        \includegraphics[width=\textwidth]{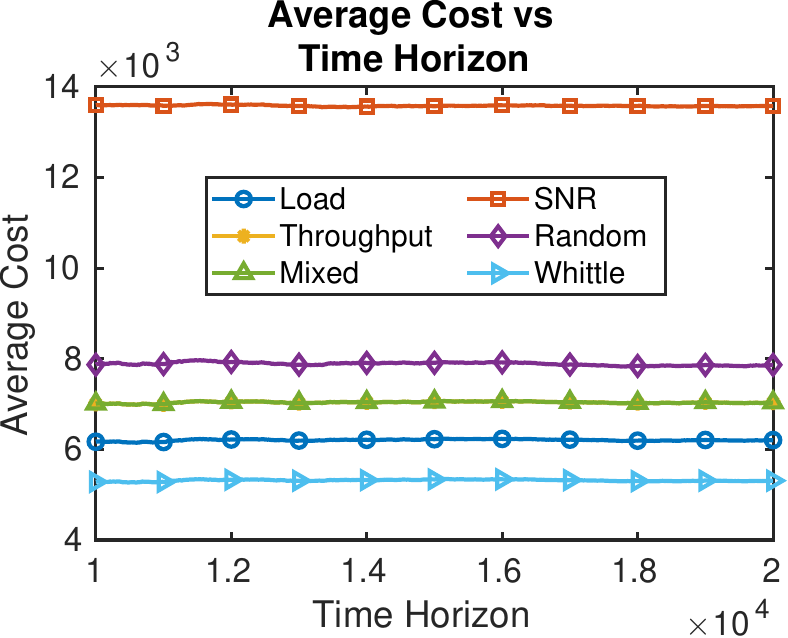}
        \caption{ }
        \label{fig:3a}
    \end{subfigure}
    \begin{subfigure}[b]{0.24\textwidth}
        \includegraphics[width=\textwidth]{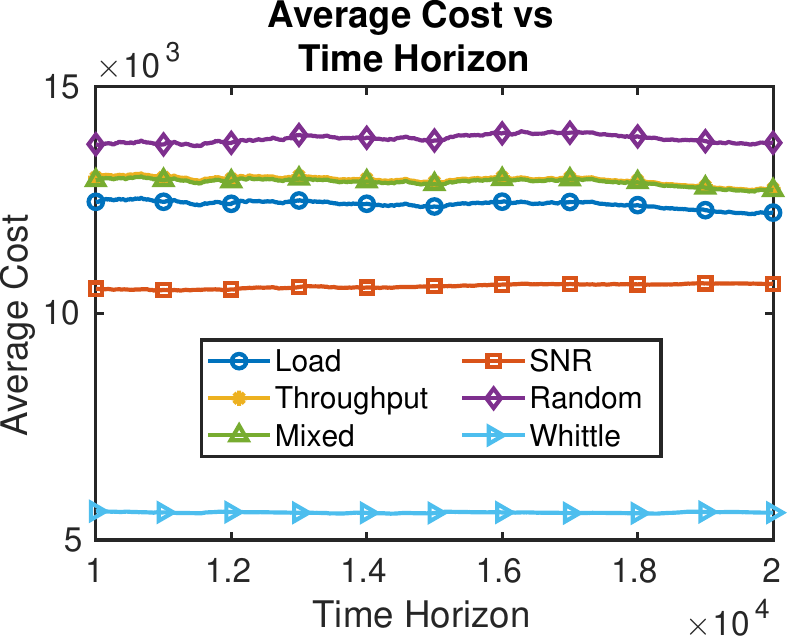}
        \caption{ }
        \label{fig:3b}
    \end{subfigure}
    \captionsetup{justification   = justified,font=scriptsize}
    \caption{The figures compare the average costs achieved under the six association policies. The following parameter values are used for both the plots: $K=8$, $r=[0.78, 0.70, 0.65, 0.60, 0.56, 0.50, 0.48, 0.45]$. The following parameter values are used for figure (a): $M=100$, $L=20$, $p_0 = 0.4$, $p_j = 0.006,\; \forall j \in \{1,2,\ldots,M\}$, $C=[95, 80, 72, 65, 58, 47, 40, 32]$. The following parameter values are used for figure (b): $M=150$, $L=10$, $p_0 = 0.7$ $p_j = 0.002,\; \forall j \in \{1,2,\ldots,M\}$,  $C=[85, 75, 68, 63, 57, 49, 45, 36]$.}
    \label{fig:3}
\end{figure}

\begin{figure}[H]
    \centering
    \begin{subfigure}[b]{0.238\textwidth}
        \includegraphics[width=\textwidth]{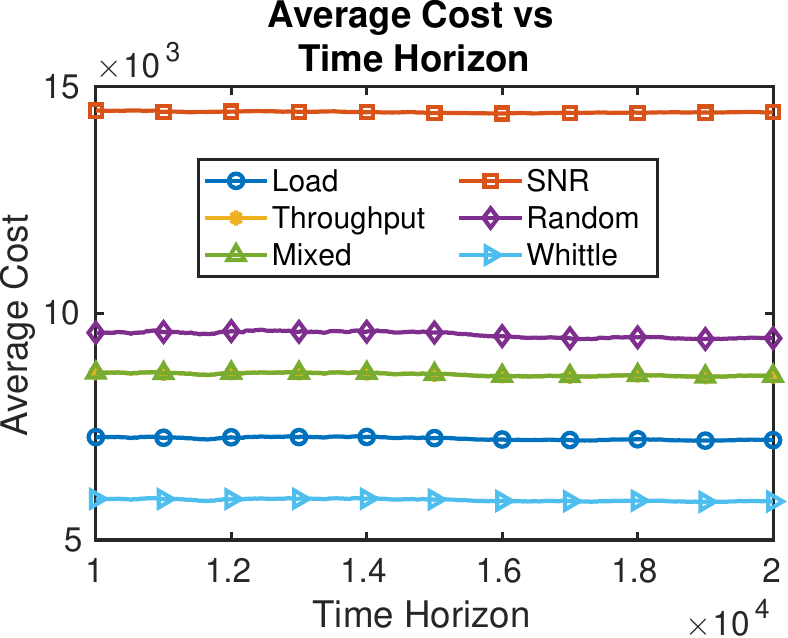}
        \caption{ }
        \label{fig:4a}
    \end{subfigure}
    \begin{subfigure}[b]{0.241\textwidth}
        \includegraphics[width=\textwidth]{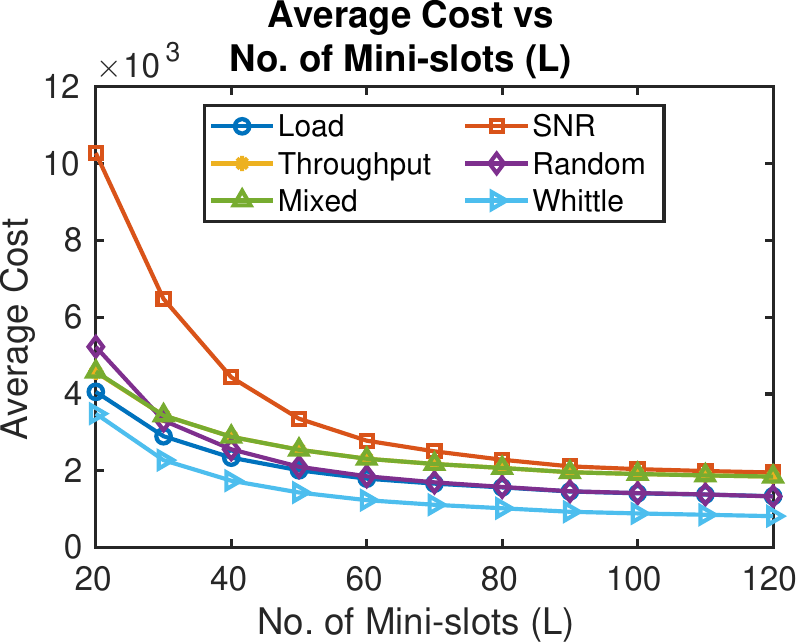}
        \caption{ }
        \label{fig:4b}
    \end{subfigure}
    \captionsetup{justification   = justified,font=scriptsize}
    \caption{The figures compare the average costs achieved under the six association policies. The maximum number of arrivals used for both the plots is: $M=100$. The parameter values used for figure (a) are: $K=10$, $L=15$, $p_{0} = 0.4$, $p_j = 0.006,\; \forall j \in \{1,2,\ldots,M\}$, $r=[0.78,0.75,0.70,0.65,0.58,0.52,0.48,0.46,0.44,0.42]$, $C=[95,85,75,65,58,47,40,32,28,25]$. The parameter values used for figure (b) are: $K=5$, $p_{0} = 0.6$, $p_j = 0.004,\; \forall j \in \{1,2,\ldots,M\}$, $r=[0.78,0.65,0.56,0.50,0.45]$, $C=[95,75,58,40,32]$. Also, in figure (b), $L$ varies from $20$ to $120$.}
    \label{fig:4}
\end{figure}

\begin{figure}[h]
    \centering
    \begin{subfigure}[b]{0.24\textwidth}
        \includegraphics[width=\textwidth]{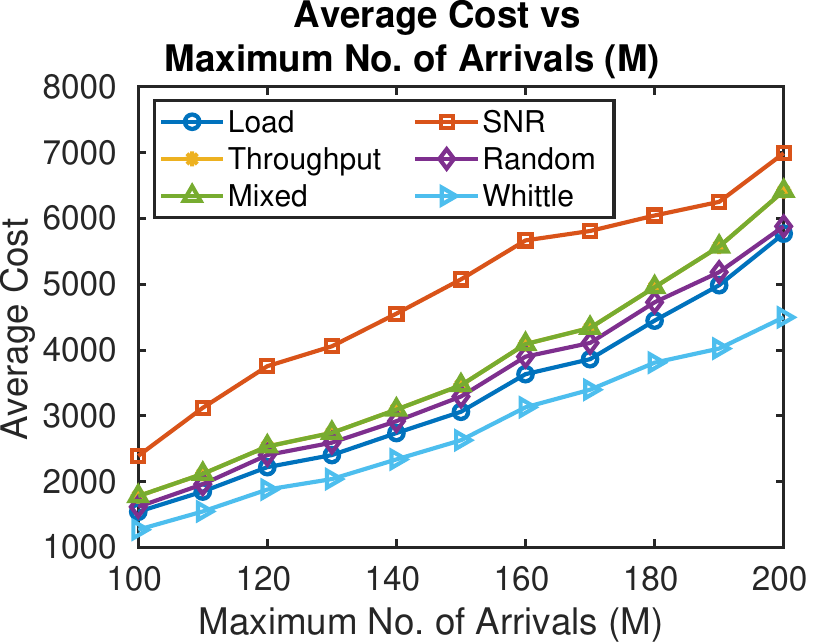}
        \caption{ }
        \label{fig:5a}
    \end{subfigure}
    \begin{subfigure}[b]{0.24\textwidth}
        \includegraphics[width=\textwidth]{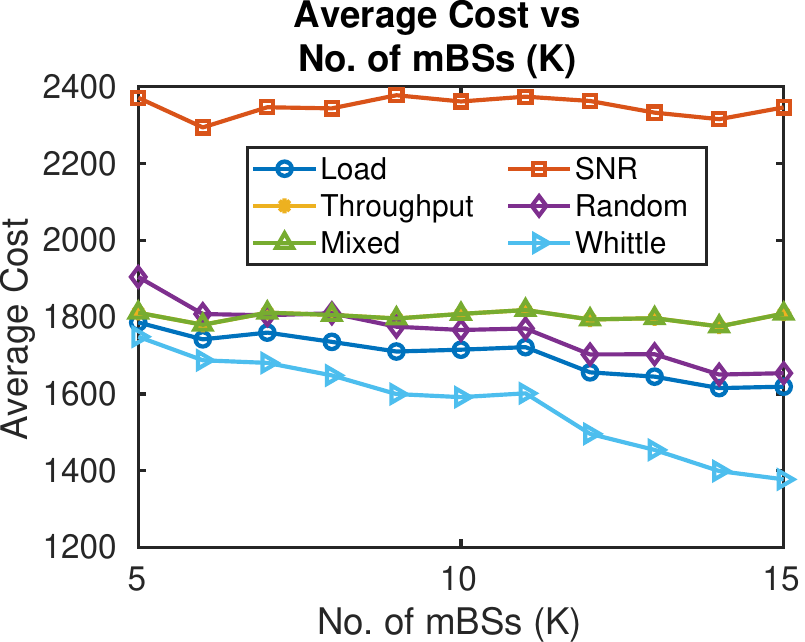}
        \caption{ }
        \label{fig:5b}
    \end{subfigure}
    \captionsetup{justification   = justified,font=scriptsize}
    \caption{The figures compare the average costs achieved under the six association policies. The values $L=30$ and $p_{0} = 0.8$ are used for both the figures. The following parameter values are used for figure (a): $K=6$, $p_j = 0.2/M,\; \forall j \in \{1,2,\ldots,M\}$, $r=[0.78,0.70,0.65,0.60,0.52,0.46]$, $C=[92,81,70,63,52,40]$, buffer size $= 250$ and different $M$ varying from $100$ to $200$. The following parameter values are used for figure (b): $M=100$, $p_j = 0.002,\; \forall j \in \{1,2,\ldots,M\}$, and different $K$ varying from $5$ to $15$. For $K=5$, the following parameter values are used: $r = [0.78,0.75,0.72,0.69,0.66]$ and $C = [90,86,82,78,74]$. For every subsequent addition of the $i^{th}$ mBS, the values of $r_i$ and $C_i$ are selected as $0.81-0.3i$ and $94-4i$, respectively, where $i \in \{6,7,\ldots,15\}$.}
    \label{fig:5}
\end{figure} 

In our simulations, the initial states of all the mBSs are set to zero. The probability that no user arrives in a slot, i.e., $p_0$, is varied and the remaining probabilities are assumed to be equal, i.e., $p_j = \frac{1-p_0}{M}, \; \forall j \in \{1,2,\ldots,M\}$. Also, the  simulations utilize a time horizon, denoted as $T$, consisting of $20,000$ slots. The performance of the six policies in terms of the average cost over the last $10,000$ slots is plotted. This choice is made to evaluate the long-term average cost rather than short-term fluctuations. Unless otherwise mentioned, the buffer size of each mBS is considered to be $200$ packets. Let $r = [r_1,\ldots, r_K]$ and $C = [C_1,\ldots,C_K]$, where $r_i$ and $C_i$ represent the serving rate and holding cost incurred at mBS $i$, respectively.

\begin{table}[ht]
    \captionsetup{justification   = justified,font=scriptsize}
    \caption{The table compares the average delays computed for different values of $K$ varying from $2$ to $10$ under the  six association policies. The following parameter values are used: $M=100$, $L=35$, $p_0 = 0.3$, and $p_j = 0.007,\; \forall j \in \{1,2,\ldots,M\}$. For $K=2$, the following parameter values are used: $r = [0.77,0.765]$ and $C = [70,69.75]$. For every subsequent addition of the $i^{th}$ mBS, the values of $r_i$ and $C_i$ are selected as $0.775-0.05i$ and $70.25-0.25i$, respectively, where $i \in \{3,4,\ldots,10\}$.}
\renewcommand{\arraystretch}{1.5}
\footnotesize
    \centering
    \scalebox{0.7}{
    \begin{tabular}{||P{0.16\linewidth}|P{0.08\linewidth}|P{0.09\linewidth}|P{0.16\linewidth}|P{0.1\linewidth}|P{0.08\linewidth}|P{0.09\linewidth}||}
    \hline \hline
         \multirow{2}{1.5cm}{\centering No. of mBSs (K)} & \multicolumn{6}{c||}{Average Delay} \\
         \cline{2-7}
          & Load & SNR & Throughput & Random & Mixed & Whittle \\
         \hline 
         2 & 42.60 & 142.26  & 42.62 & 69.46 & 42.62 & 42.60\\ \hline
         3 & 33.94 & 142.37  & 33.90 & 52.26 & 33.89 & 33.83\\ \hline
         4 & 33.80 & 144.17  & 33.65 & 47.17 & 33.65 & 33.61\\ \hline
         5 & 33.88 & 142.41  & 33.69 & 43.71 & 33.69 & 33.61\\ \hline
         6 & 34.01 & 142.79  & 33.68 & 41.42 & 33.68 & 33.61\\ \hline
         7 & 34.14 & 142.23  & 33.71 & 40.65 & 33.71 & 33.63\\ \hline
         8 & 34.23 & 143.20  & 33.68 & 39.70 & 33.68 & 33.60\\ \hline
         9 & 34.27 & 143.85  & 33.66 & 39.28 & 33.66 & 33.59\\ \hline
         10 & 34.44 & 142.67  & 33.64 & 39.16 & 33.64 & 33.56 \\ \hline \hline
         \end{tabular}}
    \label{tab:avg_delay_vs_K}
\end{table}

\begin{table}[ht]
    \captionsetup{justification   = justified,font=scriptsize}
    \caption{The table compares the average delays computed for different values of $L$ varying from $15$ to $55$ under the  six association policies. The following parameter values are used: $M=100$, $K=5$, $p_0 = 0.6$ $p_j = 0.004,\; \forall j \in \{1,2,\ldots,M\}$, $r=[0.77, 0.76, 0.75, 0.74, 0.73]$, and $C=[70, 69.5, 69, 68.5, 68]$.}
\renewcommand{\arraystretch}{1.5}
\footnotesize
    \centering
    \scalebox{0.7}{
    \begin{tabular}{||P{0.22\linewidth}|P{0.08\linewidth}|P{0.09\linewidth}|P{0.16\linewidth}|P{0.1\linewidth}|P{0.08\linewidth}|P{0.09\linewidth}||}
    \hline \hline
         \multirow{2}{1.9cm}{\centering No. of Mini-slots (L)} & \multicolumn{6}{c||}{Average Delay} \\
         \cline{2-7}
          & Load & SNR & Throughput & Random & Mixed & Whittle \\
         \hline
         15 & 34.81 & 166.20  & 34.04 & 52.52 & 34.04 & 33.96\\ \hline
         20 & 34.38 & 138.96  & 33.87 & 47.46 & 33.87 & 33.79\\ \hline
         25 & 34.31 & 110.91  & 33.77 & 43.2 & 33.77 & 33.72\\ \hline
         30 & 34.35 & 91.33  & 33.76 & 41.04 & 33.76 & 33.69\\ \hline
         35 & 34.32 & 76.63  & 33.72 & 39.27 & 33.72 & 33.59\\ \hline
         40 & 34.30 & 66.94  & 33.65 & 38.25 & 33.65 & 33.59\\ \hline
         45 & 34.27 & 59.02  & 33.66 & 37.59 & 33.66 & 33.56\\ \hline
         50 & 34.33 & 53.52  & 33.67 & 37.24 & 33.67 & 33.54\\ \hline
         55 & 34.32 & 49.13  & 33.65 & 36.54 & 33.65 & 33.57 \\  \hline \hline
         \end{tabular}}
    \label{tab:avg_delay_vs_L}
\end{table}

\begin{table}[ht]
    \captionsetup{justification   = justified,font=scriptsize}
    \caption{The table compares the average throughput computed for different values of $K$ varying from $2$ to $10$ under the  six association policies. The following parameter values are used: $M=100$, $L=35$, $p_0 = 0.3$, and $p_j = 0.007,\; \forall j \in \{1,2,\ldots,M\}$.   For $K=2$, the following parameter values are used: $r = [0.77,0.765]$ and $C = [70,69.75]$. For every subsequent addition of the $i^{th}$ mBS, the values of $r_i$ and $C_i$ are selected as $0.775-0.05i$ and $70.25-0.25i$, respectively, where $i \in \{3,4,\ldots,10\}$.}
\renewcommand{\arraystretch}{1.5}
\footnotesize
    \centering
    \scalebox{0.7}{
    \begin{tabular}{||P{0.16\linewidth}|P{0.08\linewidth}|P{0.08\linewidth}|P{0.16\linewidth}|P{0.1\linewidth}|P{0.08\linewidth}|P{0.09\linewidth}||}
    \hline \hline
         \multirow{2}{1.5cm}{\centering No. of mBSs (K)} & \multicolumn{6}{c||}{Average Throughput } \\
         \cline{2-7}
          & Load & SNR & Throughput & Random & Mixed & Whittle \\
         \hline
         2 & 44.92 & 13.96  & 44.92 & 34.53 & 44.94 & 45.09\\ \hline
         3 & 51.52 & 13.96  & 51.62 & 41.53 & 51.61 & 51.69\\ \hline
         4 & 51.70 & 13.75  & 51.89 & 44.21 & 51.89 & 51.98\\ \hline
         5 & 51.55 & 13.83  & 51.83 & 45.76 & 51.83 & 51.97\\ \hline
         6 & 51.35 & 13.81  & 51.85 & 46.89 & 51.85 & 51.97\\ \hline
         7 & 51.23 & 14.08  & 51.32 & 47.28 & 51.83 & 51.95\\ \hline
         8 & 51.08 & 13.82  & 51.87 & 47.66 & 51.87 & 52.00\\ \hline
         9 & 50.93 & 13.68  & 51.85 & 47.96 & 51.85 & 51.93\\ \hline
         10 & 50.74 & 13.90  & 51.94 & 47.91 & 51.94 & 52.08 \\  \hline \hline
         \end{tabular}}
    \label{tab:avg_thru_vs_K}
\end{table}

\begin{table}[ht]
    \captionsetup{justification   = justified,font=scriptsize}
    \caption{The table compares the Jain's fairness indices computed for different values of $K$ varying from $2$ to $10$ under the  six association policies. The following parameter values are used: $M=100$, $L=35$, $p_0 = 0.3$, and $p_j = 0.007,\; \forall j \in \{1,2,\ldots,M\}$.   For $K=2$, the following parameter values are used: $r = [0.77,0.765]$ and $C = [70,69.75]$. For every subsequent addition of the $i^{th}$ mBS, the values of $r_i$ and $C_i$ are selected as $0.775-0.05i$ and $70.25-0.25i$, respectively, where $i \in \{3,4,\ldots,10\}$.}
\renewcommand{\arraystretch}{1.5}
\footnotesize
    \centering
    \scalebox{0.7}{
    \begin{tabular}{||P{0.16\linewidth}|P{0.09\linewidth}|P{0.09\linewidth}|P{0.16\linewidth}|P{0.1\linewidth}|P{0.09\linewidth}|P{0.09\linewidth}||}
    \hline \hline
         \multirow{2}{1.5cm}{\centering No. of mBSs (K)} & \multicolumn{6}{c||}{Jain's Fairness Index} \\
         \cline{2-7}
          & Load & SNR & Throughput & Random & Mixed & Whittle \\
         \hline
         2 & 0.1217 & 0.0737  & 0.1217 & 0.1025 & 0.1217 & 0.1219\\ \hline
         3 & 0.1969 & 0.1105  & 0.1968 & 0.1721 & 0.1968 & 0.1969\\ \hline
         4 & 0.2633 & 0.1469  & 0.2634 & 0.2386 & 0.2634 & 0.2634\\ \hline
         5 & 0.3289 & 0.1853  & 0.3289 & 0.3054 & 0.3289 & 0.3292\\ \hline
         6 & 0.3948 & 0.2222  & 0.3949 & 0.3726 & 0.3949 & 0.3950\\ \hline
         7 & 0.4606 & 0.2559  & 0.4607 & 0.4375 & 0.4607 & 0.4608\\ \hline
         8 & 0.5262 & 0.2936  & 0.5266 & 0.5042 & 0.5266 & 0.5267\\ \hline
         9 & 0.5917 & 0.3317  & 0.5925 & 0.5697 & 0.5925 & 0.5925\\ \hline
         10 & 0.6576 & 0.3678  & 0.6583 & 0.6341 & 0.6583 & 0.6584\\ \hline \hline
         \end{tabular}}
    \label{tab:jfi_vs_K}
\end{table}

Figs. \ref{fig:2}-\ref{fig:5} show that the Whittle index based policy outperforms every other policy in terms of the average cost for all the parameter values considered. Also, the SNR based policy performs the worst among all the policies. This is because the SNR based policy associates users to mBSs only based on the data rates and does not perform load balancing. Tables \ref{tab:avg_delay_vs_K} and \ref{tab:avg_delay_vs_L} show that the Whittle index based policy achieves a lower average delay than all the other policies. Finally, Tables \ref{tab:avg_thru_vs_K} and \ref{tab:jfi_vs_K} show that the Whittle index based policy achieves a higher average throughput and JFI than all the other policies, respectively. In summary, our proposed Whittle index based policy outperforms all the other policies in terms of average cost, average delay, average throughput, as well as JFI for all the parameter values considered.

\section{Conclusions And Future Work}\label{Section12_con}
In this paper, we considered the problem of user association in a dense mmWave network, in which, each arriving user brings a file containing a random number of packets and each time slot is divided into multiple mini-slots. 
 We formulated this problem as a restless multi-armed bandit problem and proved that it is Whittle indexable. Based on this result, we presented a scheme for computing the Whittle indices of different mBSs, and proposed an association scheme under which, each arriving user is associated with the mBS with the smallest value of the Whittle index. Using extensive simulations, we showed that the proposed Whittle index based scheme outperforms several association schemes proposed in prior work in terms of various performance metrics such as average cost, delay, throughput, and JFI. A direction for future research is to extend the results of this paper to the case where each mBS operates over multiple channels, on which frequency-selective fading takes place.

\ifCLASSOPTIONcaptionsoff
  \newpage
\fi



\bibliographystyle{IEEEtran}
\bibliography{main}
%

%









\end{document}